\documentclass{ws-ijmpd}

\usepackage{amssymb}
\usepackage{epsfig}
\usepackage{latexsym}

\def\sqr#1#2{{\vcenter{\hrule height.#2pt\hbox{\vrule width.#2pt
height#1pt \kern#1pt \vrule width.#2pt}\hrule height.#2pt}}}

\def\negenspace{\kern-1.1em}

\def\a{\alpha}
\def\b{\beta}
\def\g{\gamma}

\def\vta{\vartheta}

\def\qslash{{\not\!Q}}
\def\zslash{{\not\!Z}}

\def\D0{\stackrel{\circ}{D}}

\begin{document}

%
\def\nocropmarks{\vskip5pt\phantom{cropmarks}}
%

\markboth{Peter Baekler and Friedrich W.\ Hehl}
{Rotating black holes in metric-affine gravity}

%
\catchline{}{}{}
%

\title{ROTATING BLACK HOLES IN METRIC-AFFINE GRAVITY
}
\author{\footnotesize PETER
  BAEKLER\footnote{peter.baekler@fh-duesseldorf.de,
    http://www.et.fh-duesseldorf.de/home/baekler/index.html}}

\address{Fachbereich Medien, Fachhochschule D\"usseldorf\\ University
  of Applied Sciences\\ Josef-Gockeln-Str.\ 9\\40474 D\"usseldorf,
  Germany
}
\author{FRIEDRICH W.\ HEHL\footnote{hehl@thp.uni-koeln.de,
    http://www.thp.uni-koeln.de/gravitation/}}

\address{Institute for Theoretical Physics\\ University of Cologne\\ 50923
  K\"oln, Germany\\ and\\ Department of Physics and Astronomy\\
  University of Missouri-Columbia\\ Columbia, MO 65211, USA}

\maketitle


\begin{abstract}
  Within the framework of metric-affine gravity (MAG, metric and an
  independent linear connection constitute spacetime), we find, for a
  specific gravitational Lagrangian and by using {\it prolongation\/}
  techniques, a stationary axially symmetric exact solution of the
  vacuum field equations. This black hole solution embodies a
  Kerr-deSitter metric and the post-Riemannian structures of torsion
  and nonmetricity. The solution is characterized by mass, angular
  momentum, and shear charge, the latter of which is a measure for
  violating Lorentz invariance. {\it file Baekler/Magaxi13.tex}
  \medskip

\medskip

\end{abstract}

\section{Introduction}

In the spirit of a gauge-theoretical approach to gravity, a
metric-affine gauge field theory of gravitation (``metric-affine
gravity'' MAG) has been proposed\footnote{Reviews have been provided
  in Ref.\cite{PRs} and Ref.\cite{Erice95}. The group of Dereli,
  Tucker, and Wang\cite{Dereli1,TuckerWang1,TuckerWang2} applied such
  theories to the dark matter problem, inter alia, Minkevich et
  al.\cite{Minkevich95,MinkevichVas2003,Minkevich2005z,Minkevich2005a,Minkevich2005b},
  Puetzfeld et
  al.\cite{Puetzfeld2001,Puetzfeld2002,Puetzfeld2004a,Puetzfeld2004b},
  and Babourova \&
  Frolov\cite{Babourova2003,Babourova2004,Babourova2005} mainly to
  cosmological solutions. New exact solutions were found, amongst
  others, by Vassiliev \& King\cite{King,Dima1,Dima2,Pasic}. The
  determination of the energy and of the other conserved quantities of
  exact solutions of MAG has been developed in particular by Nester
  and his group\cite{Chen94,Yu-Huei1,Nester2000,Yu-Huei2,Yu-Huei3},
  see also Baekler et al.\cite{BaeklerShirafuji}. Comparison with
  observations have been pioneered by Preuss\cite{Preuss2004} and
  Solanki\cite{Solanki2004}, see also Puetzfeld, loc.cit..}  based on
the metric $g$ and the affine group $A(4,R)$, i.e., the semi-direct
product of the four dimensional translational group $R^{4}$ and the
general linear group $GL(4,R)$.  Besides the usual ``weak''
Newton-Einstein type gravity, additional ``strong gravity'' pieces
will arise that are supposed to be mediated by additional geometrical
degrees of freedom related to the independent linear connection 1-form
$\Gamma_\a{}^\b=\Gamma_{i\a}{}^\b dx^i$. Here $\a,\b,\dots =0,1,2,3$
denote frame (or anholonomic) indices and $i,j,\dots =0,1,2,3$
coordinate (or holonomic) indices.  Alternatively, the strong gravity
pieces can also be expressed in terms of the nonmetricty 1-form
$Q^{\alpha\beta}=Q_{i}{}^{\alpha\beta}dx^{i}$ and the torsion 2-form
${T^{\alpha}=\frac{1}{2}T_{ij}{}^{\alpha} dx^{i}\wedge dx^{j}}$. The
propagating modes related to the new degrees of freedom manifest
themselves in post-Riemannian pieces of the curvature $R_\a{}^\b$.

\section{Geometrical structures of a metric-affine spacetime}

We briefly summarize the basic notions of metric-affine geometry.  Let
us start from a n-dimensional differentiable manifold $M_{n}$. At each
point $P\in M_{n}$ we can construct the n-dimensional tangent vector
space $T_{P}(M_{n})$ with vector basis (or {\it frame}) $e_{\alpha}$.
In the space $T_{P}^{\ast}(M_{n})$, dual to $T_{P}(M_{n})$, we
introduce a local one-form basis (or {\it coframe})
${\vartheta}^{\alpha}$ such that
\begin{equation}
  e_{\alpha}\rfloor {\vartheta}^{\beta} =
  {\delta}_{\alpha}^{\beta}\,,
\end{equation}
where $\rfloor $ symbolizes the interior product. Generally, the coframe
is not integrable, i.e., we have
\begin{equation}
  C^\a:= d{\vartheta}^{\alpha} =
  \frac{1}{2}{C}_{\mu\nu}{}^{\alpha}{\vartheta}^{\mu}\wedge
  {\vartheta}^{\nu} \neq 0 \, ,
\end{equation}
where the 1-form $C^{\alpha}$ is a measure of the anholonomity.

We assume that the manifold is endowed with a {\it metric\/} $g$. We
decompose it with respect to the coframe ${\vartheta}^{\alpha}$ and
find
\begin{equation}
  g = g_{\alpha\beta}{\vartheta}^{\alpha}\otimes
  {\vartheta}^{\beta}\,.
\end{equation}
Furthermore, we assume that the manifold $M_{n}$ carries additionally
a metric-inde\-pendent {\it linear connection\/}
${\Gamma}_{\alpha}{}^{\beta}$.  Accordingly, nonmetricity and torsion
emerge as {\em geometrical field strengths}, to be defined as
\begin{equation}
Q_{\alpha\beta} := -Dg_{\alpha\beta}
\end{equation}
and
\begin{equation}
  T^{\alpha} := d{\vartheta}^{\alpha}+ {\Gamma}_{\mu}{}^{\alpha}\wedge
  {\vartheta}^{\mu} = D{\vartheta}^{\alpha}\,,
\end{equation}
respectively, together with the curvature 2-form
\begin{equation}
  R_{\alpha}{}^{\beta} := d{\Gamma}_{\alpha}{}^{\beta}
  -{\Gamma}_{\alpha}{}^{\mu}\wedge {\Gamma}_{\mu}{}^{\beta}\,.
\end{equation}
Here $D$ is the exterior covariant derivative with respect to the
connection $\Gamma_\a{}^\b$.

The geometrical field strengths give rise to integrability conditions,
namely to {\it Bianchi} and {\it Ricci} identities. We have,
with\footnote{Parentheses surrounding indices $(\a\b) :=(\a\b+\b\a)/2$
  denote symmetrization and brackets $[\a\b] :=(\a\b-\b\a)/2$
  antisymmetrization.} $Z_{\a\b}:=R_{(\a\b)}$,
\begin{eqnarray}
  DDg_{\alpha\beta} & = & -DQ_{\alpha\beta} =
  -2R_{(\alpha}{}^{\mu}g_{\beta )\mu}=-2Z_{\a\b}\, , \\ \label{1stBia}
  DD{\vartheta}^{\alpha} & = & DT^{\alpha} = R_{\mu}{}^{\alpha}\wedge
  {\vartheta}^{\mu}\, , \\ DR_{\alpha}{}^{\beta} & = & 0 \, ;\\
  DDT^{\alpha} & = & \left(DR_{\mu}{}^{\alpha}\right)\wedge
  {\vartheta}^{\mu} + R_{\mu}{}^{\alpha}\wedge T^{\mu} =
  R_{\mu}{}^{\alpha}\wedge T^{\mu}\, , \\ DDQ_{\alpha\beta} & = &
  -2R_{(\alpha}{}^{\mu}\wedge Q_{\beta )\mu}=:S_{\alpha\beta}\, , \\
  DS_{\alpha\beta} & = & DDDQ_{\alpha\beta} =
  -4R_{(\alpha}{}^{\mu}\wedge Z_{\beta )\mu}\,.
\end{eqnarray}

We can, in terms of the metric field of MAG, always construct a
Riemannian (or Levi-Civita) connection.  Therefore, for the purpose of
a comparison with general relativity, e.g., it is useful to decompose
the connection $\Gamma_{\a\b}:=\Gamma_\a{}^\mu g_{\b\mu}$ into a
Riemannian piece $\widetilde{\Gamma}_{\alpha\beta} $ and a tensorial
post-Riemannian piece $N_{\alpha\beta}$,
\begin{equation}\label{connection}
  {\Gamma}_{\alpha\beta} = \widetilde{ \Gamma}_{\alpha\beta}+
  N_{\alpha\beta}\,.
\end{equation}
The distortion 1-form $N_{\alpha\beta}$ allows us to recover nonmetricity
and torsion according to
\begin{equation}\label{distortion}
  Q_{\alpha\beta}=2N_{(\alpha\beta)}\,,\qquad
  T^\alpha=N_\beta{}^\alpha\wedge\vartheta^\beta\,.
\end{equation}
Explicitly, we have
\begin{equation}
  N_{\alpha\beta}=-e_{[\alpha}\rfloor T_{\beta]} + {1\over
    2}(e_{\alpha}\rfloor e_{\beta}\rfloor
  T_{\gamma})\,\vartheta^{\gamma} + (e_{[\alpha}\rfloor
  Q_{\beta]\gamma})\,\vartheta^{\gamma} +{1\over
    2}Q_{\alpha\beta}\,,\label{N}
\end{equation}
see Ref.\cite{PRs}, Eq.(3.10.7).

In a metric-affine spacetime we can separate the curvature 2-form
$R_{\alpha\beta}$ into a tracefree symmetric part (``shear'')
${\zslash}_{\alpha\beta}$, a trace part (``dila[ta]tion'') $Z$, and an
antisymmetric part (``rotation'') $W_{\a\b}$ according to
\begin{equation}
R_{\alpha\beta} = R_{( \alpha\beta )} + R_{[ \alpha\beta ]} =
{\zslash}_{\alpha\beta}+\frac{1}{4}Zg_{\alpha\beta}+W_{\alpha\beta}\,,
\end{equation}
with the definitions
\begin{equation}\label{curvdef}
  Z_{\a\b}:=R_{(\a\b)},\quad \zslash_{\a\b}:=Z_{\a\b}-\frac
  14\,Zg_{\a\b},\quad Z:=Z_\a{}^\a,\quad W_{\a\b}:=R_{[\a\b]}.
\end{equation}
The symmetric part $Z_{\alpha\beta}$ represents the
post-Riemann-Cartan part of the curvature, that is, it vanishes
together with $Q_{\a\b}$, whereas $W_{\alpha\beta}$ includes the
Riemannian contributions, inter alia.

We now specialize to 4-dimensional spacetime with Lorentz signature
$(-++\,+)$. Quite generally, nonmetricity $Q_{\alpha\beta}$, torsion
$T^{\alpha}$, and curvature $R_{\alpha\beta}$ can then be split into
smaller pieces, they can be decomposed irreducibly under the Lorentz
group, see \ref{appdec}. In the following table, we will give, for
$n=4$, an overview of the number of independent components of these
quantities:\bigskip

\centerline{{\it Table 1.} Number of components of the irreducible
pieces.}

$$
{\begin{tabular}{|c||c|c|c|c|c|c||c|} \hline $Q_{\alpha\beta}$ &
$^{(1)}Q_{\alpha\beta}$ & $^{(2)}Q_{\alpha\beta}$ &
$^{(3)}Q_{\alpha\beta}$ & $^{(4)}Q_{\alpha\beta}$ & - & - &
$Q_{\alpha\beta}$ \\
\hline\hline
 & 16 & 16 & 4 & 4 & - & - & $\Sigma = 40$ \\ \hline
\hline $T^{\alpha}$ & $^{(1)}T^{\alpha}$ & $^{(2)}T^{\alpha}$ &
$^{(3)}T^{\alpha}$ & - & - & - &
$T^{\alpha}$ \\
\hline\hline
 & 16 & 4 & 4 & - & - & - & $\Sigma = 24$ \\ \hline
\hline $R_{[\alpha\beta ]}$ & $^{(1)}W_{\alpha\beta}$ &
$^{(2)}W_{\alpha\beta}$ & $^{(3)}W_{\alpha\beta}$ &
$^{(4)}W_{\alpha\beta}$ & $^{(5)}W_{\alpha\beta}$ &
$^{(6)}W_{\alpha\beta}$ &
$W_{\alpha\beta}$ \\
\hline\hline
 & 10 & 9 & 1 & 9 & 6 & 1 & $\Sigma = 36$ \\ \hline
\hline $R_{(\alpha\beta )}$ & $^{(1)}Z_{\alpha\beta}$ &
$^{(2)}Z_{\alpha\beta}$ & $^{(3)}Z_{\alpha\beta}$ &
$^{(4)}Z_{\alpha\beta}$ & $^{(5)}Z_{\alpha\beta}$ & - &
$Z_{\alpha\beta}$ \\
\hline\hline
 & 30 & 9 & 6 & 6 & 9 & - & $\Sigma = 60$ \\ \hline
\hline
\end{tabular}}
$$
\bigskip

The exterior products of the coframe $\vta^\a$ are denoted by
$\vta^{\a\b}:=\vta^\a\wedge\vta^\b$, etc.. Since a metric is
prescribed, we can define a Hodge star operator $^\star $ which maps,
in 4 dimensions, $p$-forms into $(4-p)$-forms. Then, we can introduce
the eta-basis $\eta:=\,^\star 1$, $\>\eta^\a:=\,^\star\vta^\a$,
$\>\eta^{\a\b}:=\,^\star\vta^{\a\b}$, etc..

\section{Lagrangian and field equations}

We consider in the first order Lagrangian formalism the geometrical
variables $\{g_{\alpha\beta}\, , {\vartheta}^{\alpha}\, ,
{\Gamma}_{\alpha}{}^{\beta}\} $ to be minimally coupled to matter
fields, collectively called ${\Psi}$, such that the total Lagrangian,
i.e., the geometrical part plus the matter part, reads
\begin{equation}
  L_{{\rm tot}} = V(g_{\alpha\beta}\, , {\vartheta}^{\alpha}\, ,
  Q_{\alpha\beta}\, , T^{\alpha}\, ,R_{\alpha}{}^{\beta} ) +
  L_{{\rm matter}}(g_{\alpha\beta}\, , {\vartheta}^{\alpha}\, ,
  {\Psi}\, , D{\Psi})\, .
\end{equation}
By using the excitations as place holders,
\begin{equation}
M^{\alpha\beta} =
-2\frac{{\partial}V}{{\partial}Q_{\alpha\beta}}\, , \quad
H_{\alpha} = -\frac{{\partial}V}{{\partial}T^{\alpha}}\, , \quad
H^{\alpha}{}_{\beta} =
-\frac{{\partial}V}{{\partial}R_{\alpha}{}^{\beta}}\,,
\end{equation}
the field equations of metric-affine gravity can be given in a very
concise form:\cite{PRs}
\begin{eqnarray}\label{zeroth}
  DM^{\alpha\beta} - m^{\alpha\beta} &=& \sigma^{\alpha\beta}
  \qquad\,\qquad {({\delta}/{\delta}g_{\alpha\beta})}\,,\\ DH_{\alpha}
  - E_{\alpha}\hspace{6pt} & =& \Sigma_{\alpha} \qquad\>\;\qquad {
    ({\delta}/{\delta}{\vartheta}^{\alpha})}\,,\label{first}\\
  DH^{\alpha} {}_{\beta} - E^{\alpha}{}_{\beta} &=&
  \Delta^{\alpha}{}_{\beta} \qquad\qquad{
    ({\delta}/{\delta}{\Gamma}_{\alpha}{}^{\beta})}\,,\label{second}\\
  {{\delta L}\over{\delta\Psi}} &=& 0\quad\qquad\qquad\,\; {\rm
    (matter)}\,.\label{matter}
\end{eqnarray}

On the right-hand-sides of each of the three gauge field equations
(\ref{zeroth}) to (\ref{second}), there act the material currents as
sources, on the left-hand-side there are typical Yang-Mills like terms
governing the gauge fields, their first derivatives, and the
corresponding gauge field currents. The gauge currents turn out to be
the metrical (Hilbert) energy-momentum of the gauge fields
\begin{equation}\label{zerothx}\! m^{\alpha\beta}\! :=\!
  2{{\partial V}\over{\partial g_{\alpha\beta}}}=
  \vartheta^{(\alpha} \wedge E^{\beta )}+ Q^{(\beta}{}_{\gamma}\wedge
  M^{\alpha )\gamma}-T^{(\alpha}\wedge H^{\beta )} - R_\gamma {}^{(
    \alpha} \wedge H^{\gamma | \beta )} + R^{( \beta |\gamma} \wedge
  H^{\alpha )} {}_\gamma ,
\end{equation}
the canonical (Noether) energy-momentum of the gauge fields
\begin{equation}\label{firstx} E_{\alpha} :=
  {{\partial V}\over{\partial\vartheta^{\alpha}}}= e_{\alpha}\rfloor V
  + (e_{\alpha}\rfloor T^{\beta})\wedge H_{\beta} + (e_{\alpha}\rfloor
  R_{\beta}{}^{\gamma})\wedge H^{\beta}{}_{\gamma} +
  {1\over2}(e_{\alpha}\rfloor Q_{\beta\gamma})\, M^{\beta\gamma}\,,
\end{equation}
and the hypermomentum of the gauge fields
\begin{equation}\label{secondx}
  E^{\alpha}{}_{\beta}:= {{\partial
      V}\over{\partial\Gamma_{\alpha}{}^{\beta}}}= -
  \vartheta^{\alpha}\wedge H_{\beta} - g_{\beta\gamma}\,
  M^{\alpha\gamma}\,,
\end{equation}
respectively.

The {\it most general\/} parity conserving {\it quadratic\/}
Lagrangian, which is expressed in terms of the $4+ 3+ 6+5$ irreducible
pieces of $Q_{\alpha\beta}$, $T^\alpha$, $W_\alpha{}^\beta$, and
$Z_\alpha{}^\beta$, respectively, reads
\begin{eqnarray}
\label{QMA}&& V_{\rm MAG}=
\frac{1}{2{\kappa}}\,\left[-a_0\,R^{\alpha\beta}\wedge\eta_{\alpha\beta}
  -2\lambda_{0}\,\eta\right.\cr &&\hspace{20pt}\left.
  +T^\alpha\wedge{}^\star \!\left(\sum_{I=1}^{3}a_{I}\,^{(I)}
    T_\alpha\right) + Q_{\alpha\beta} \wedge{}^\star
  \!\left(\sum_{I=1}^{4}b_{I}\,^{(I)}Q^{\alpha\beta}\right)\right.
\nonumber \\ &&\hspace{20pt}\left.+
  2\left(\sum_{I=2}^{4}c_{I}\,^{(I)}Q_{\alpha\beta}\right)
  \wedge\vartheta^\alpha\wedge{}^\star \!\, T^\beta + b_{5}
  \left(^{(3)}Q_{\alpha\gamma}\wedge\vartheta^\alpha\right)\wedge
  {}^\star \!\left(^{(4)}Q^{\beta\gamma}\wedge\vartheta_\beta
  \right)\right]\\ & &\hspace{20pt} -\frac{1}{2\rho}\,R^{\alpha\beta}
\wedge{}^\star \!  \left(\sum_{I=1}^{6}w_{I}\,^{(I)}W_{\alpha\beta} +
  \sum_{I=1}^{5}{z}_{I}\,^{(I)}Z_{\alpha\beta} \nonumber\right.\\
&&\hspace{20pt} \left.  +w_7\,\vartheta_\alpha\wedge(e_\gamma\rfloor
  ^{(5)}W^\gamma{}_{\beta} ) +z_6\,\vartheta_\gamma\wedge
  (e_\alpha\rfloor ^{(2)}Z^\gamma{}_{\beta}
  )+\sum_{I=7}^{9}z_I\,\vartheta_\alpha\wedge(e_\gamma\rfloor
  ^{(I-4)}Z^\gamma{}_{\beta} )\right).\nonumber
\end{eqnarray}
see Refs.\cite{Esser,YuriEffective,MAGII,HBH_2005} and the literature
quoted there. Here $\kappa$ is the dimensionful ``weak'' gravitational
constant, $\lambda_{0}$ the ``bare'' cosmological constant, and $\rho$
the dimensionless ``strong'' gravity coupling constant. The constants
$ a_0, \ldots a_3$, $b_1, \ldots b_5$, $c_2, c_3,c_4$, $w_1, \ldots
w_7$, $z_1, \ldots z_9$ are dimensionless and are expected to be of
order unity. The constant $ a_0$ can only have the values 1 or 0
depending on whether a Hilbert-Einstein term is present or not.

We ordered the Lagrangian (\ref{QMA}) in the following way: In the
first line, we have the linear pieces, a Hilbert-Einstein type term
and the cosmological term. Some algebra yields
$R^{\alpha\beta}\wedge\eta_{\alpha\beta}= {}^{(6)}W^{\alpha\beta}
\wedge\eta_{\alpha\beta}$, that is, only the curvature scalar is left
over, as expected. In the second line, we have the pure Yang-Mills
type terms for torsion and nonmetricity. If we expand them, we find $
\sim a_1{}^{(1)}T^\alpha \wedge{}^{\star (1)}T_\alpha+\dots +
b_1{}^{(1)} Q^{\a\b}\wedge {}^{\star(1)} Q_{\a\b}+\dots$ . For a
Yang-Mills field strength $F$ we have always the Lagrangian $\sim
F\wedge{}^\star F$. In our case, for $T^\a$ and $Q^{\a\b}$, the field
strength are reducible and we can put open weighting factors in front
of each square piece. Nevertheless, the second line is the obvious
analog of a Yang-Mills Lagrangian for $T^\a$ and $Q^{\a\b}$.  In the
third line, we have ``interactions'' between $Q_{\a\b}$ and $T^\a$ and
between different irreducible pieces of $Q_{\a\b}$. In the fourth
line, we have the pure Yang-Mills terms for the rotational and the
strain curvature $\sim w_1{}^{(1)}W^{\a\b}\wedge{}
^{\star(1)}W_{\a\b}+\dots +z_1{}^{(1)}Z^{\a\b}\wedge{}
^{\star(1)}Z_{\a\b}+\dots$, and, eventually, in the last line,
``exotic'' interactions between different irreducible pieces of the
curvature enter that we will drop subsequently. In other words, we
restrict ourselves to
\begin{equation}\label{exotic}
  w_7=z_6=z_7 =z_8 =z_9=0\,.
\end{equation}

Taking into consideration (\ref{exotic}), the various excitations
$\{M^{\alpha\beta}\,,H_{\alpha}\,, H^{\alpha}{}_{\beta}\}$ are found
to be
\begin{eqnarray}\label{M-excit}
  M^{\alpha\beta} & = & -\frac{2}{\kappa} ^{\star }\! \left(
    \sum\limits_{I=1}^{4} b_{I} {^{(I)}Q}^{\alpha\b} \right) \cr & & \cr
  & & -\frac{2}{\kappa}\left[ c_{2}{\vartheta}^{({\alpha}}\wedge {}
    ^{\star (1)}T^{\beta )} + c_{3}{\vartheta}^{({\alpha}}\wedge {}
    ^{\star (2)}T^{\beta )} + \frac{1}{4}(c_{3}-c_{4})\,^{\star
      }Tg^{\alpha\beta} \right] \cr & & \cr & &
  -\frac{b_{5}}{\kappa}\left[ {\vartheta}^{( \alpha}\wedge
    {}^{\star}(Q\wedge {\vartheta}^{\beta )})-\frac{1}{4}
    g^{\alpha\beta}{}\,^{\star}(3Q+{\Lambda})\right]\,,\\&&\cr
  H_{\alpha}& =& -\frac{1}{\kappa} ^{\star }\!\left(
    \sum\limits_{I=1}^{3} a_{I} {^{(I)}T}_{\alpha} +
    \sum\limits_{K=2}^{4} c_{K} {^{(K)}Q}_{\alpha\beta}\wedge
    {\vartheta}^{\beta}\right)\,,
\label{Ha-excit}\\ &&\cr\label{Hab-excit}
H^{\alpha}{}_{\beta}&=&\frac{a_{0}}{2\kappa}{\eta}^{\alpha}{}_{\beta}+
\sum\limits_{I=1}^{6}w_{I}{}^{\star (I)}W^{\alpha}{}_{\beta} +
\sum\limits_{K=1}^{5}z_{K}{}^{\star (K)}Z^{\alpha}{}_{\beta}\, .
\end{eqnarray}
The last equation can be slightly rewritten as
\begin{equation}\label{H_ab}
  H^{\alpha}{}_{\beta}=\left(\frac{a_{0}}{2\kappa}- \frac{w_{6}}{12\rho}\,W
  \right){\eta}^{\alpha}{}_{\beta}+\frac{1}{\rho} \sum\limits_{n=1}^{5} \left(
    w_{n}\, ^{\star (n)}W^{\alpha}{}_{\beta} + z_{n}\, ^{\star
      (n)}Z^{\alpha}{}_{\beta}\right)\, ,
\end{equation}
where $^{(6)}W^{\alpha\beta}=-W{\vartheta}^{\alpha\beta}/12$
corresponds to the curvature scalar $W$.

\section{Master equation for solving the field equations algebraically}

Generally, it is a very delicate task to solve the nonlinear partial
differential equations (\ref{zeroth}) to (\ref{second}) for the set of
variables $\{g_{\a\b}\,,\vartheta^{\alpha}\, , T^{\alpha}\, ,
Q_{\alpha\beta}\}$. Even for high symmetries, there will be very few
chances to find exact solutions. Therefore, we developed an algebraic
method for solving the field equations.

The main observation is that we can construct an algebraic relation
between torsion and nonmetricity. This is a result of {\em
  Prolongation Theory} that has been applied very successfully in the
context of Einstein's field equation by M.G\"urses\cite{Guerses_a}
and Bilge et al.\cite{Bilge_1986}, amongst others.  Application of
this method to the Poincar\'e gauge field theory, i.e., to MAG with
vanishing nonmetricity, $Q_{\a\b}=0$, leads to the construction of
stationary axisymmetric solutions with dynamic torsion, see Baekler et
al.\cite{BaeGurses_1987,McKerr,BaeKerr,BaeProlongation}. This method
has been developed further systematically and, in ref.\cite{MAGIII},
we used a quite general ansatz for solving also the field equations of
MAG.

It has been shown\cite{Baekler2003,MAGIII} that the following {\it
  linear relationship between nonmetricity and torsion} can be
exploited for solving the field equations of MAG straightforwardly:
\begin{equation}
T^{\alpha} =
\sum\limits_{A=2}^{4}{\tilde{\xi}}_{A}{}{}^{(A)}Q^{\alpha}{}_{\mu}\wedge
{\vartheta}^{\mu}+ {}^{(3)}T^{\alpha}\,. \label{T_general}
\end{equation}
The parameters ${\tilde{\xi}}_{A}$ have to be determined by the field
equations.

To demonstrate the consequences of such an ansatz, we will consider a
{\it simplified\/} version of (\ref{T_general}) in the form of
\begin{equation}\label{master_1}
\mathbf{ T^{\alpha} = {\xi}_{0}Q^{\alpha}{}_{\mu}\wedge {\vartheta}^{\mu} +
{\xi}_{1}Q\wedge {\vartheta}^{\alpha} + {}^{(3)}T^{\alpha}\,.}
\end{equation}
We name this equation {\it master equation.} The constants
$\xi_0$ and $\xi_1$ will be picked later in the context of solving the
field equations.

Alternatively, we can write it as
\begin{equation}
 T^{\alpha}=
{\xi}_{0}{\qslash}{}{}^{\alpha}{}_{\beta}\wedge
{\vartheta}^{\beta} + ({\xi}_{0}+{\xi}_{1})Q\wedge
{\vartheta}^{\alpha} + {}^{(3)}T^{\alpha}\,.
\label{master_1a}
\end{equation}
The {\it Weyl covector\/} $Q$ and the {\it traceless nonmetricity\/}
$\qslash^\a{}_\b$ are defined by
\begin{equation}
  Q:=\frac 14\,Q^\a{}_\a\,,\qquad
  \qslash^\a{}_\b:=Q^\a{}_\b - Q\delta^\a_\b\,.
\end{equation}
If we make use of the 2-form $P^{\alpha}$ of (\ref{P_alpha}) and the
1-form $\Lambda$ of (\ref{Lambda}), eq.(\ref{master_1a}) translates
into\footnote{We could slightly generalize this expression for the
  torsion to $$
  T^{\alpha} =
  {\tilde\xi}_{0}P^{\alpha}+{\tilde\xi}_{1}{\Lambda}\wedge
  {\vartheta}^{\alpha}+{\tilde\xi}_{2}Q\wedge{\vartheta}^{\alpha} +
  {}^{(3)}T^{\alpha}\,, $$
  with suitable constants ${\tilde\xi}_{0}$, ${\tilde\xi}_{1}$, and
  ${\tilde\xi}_{2}$.  }
\begin{equation}
T^{\alpha} = {\xi}_{0}P^{\alpha} -
\frac{{\xi}_{0}}{3}{\Lambda}\wedge {\vartheta}^{\alpha} +
({\xi}_{0}+{\xi}_{1})Q\wedge {\vartheta}^{\alpha} +
{}^{(3)}T^{\alpha} \label{master_2}\,.
\end{equation}
We compute the trace of this equation by contracting it with the frame
$e_\alpha$. Since $e_\alpha\rfloor P^\alpha=0$ and
$e_\alpha\rfloor{}^{(3)}T^\alpha=0$, we find
\begin{equation}\label{traceansatz}
  T = {\xi}_{0}\,{\Lambda} - 3({\xi}_{0}+{\xi}_{1})\,Q \,,
\end{equation}
with the 1-form $T:=e_\alpha\rfloor T^\alpha$.

Empirically, a special case of relation (\ref{traceansatz}) has been
used in MAG for constructing exact solutions in the form of the
{\it triplet ansatz}\cite{OVETH,Vlachynsky_1996,YuriEffective,GarciaMAG}
\begin{equation}\label{triplet}
  Q/k_0=\Lambda/k_1=T/k_2\,,
\end{equation}
with some constants $k_0,k_1,k_2$. We refere here to the triplet of
1-forms $Q,\Lambda,T$. Spherically symmetric solutions\cite{OVETH}
were found as well as stationary axially symmetric
ones.\cite{Vlachynsky_1996,YuriEffective,GarciaMAG} A deeper
understanding of why this ansatz (\ref{traceansatz}) could work
successfully in those approaches has been elaborated systematically by
Baekler et al.\cite{MAGIII}, see also Heinicke et al.\cite{HBH_2005},
demonstrating that the key is to look for further integrability
conditions.  Especially the first Bianchi identity (\ref{1stBia})
turns out to be helpful in answering this question.

We turn now to the connection and thus to the distortion 1-form
$N_{\a\b}$. We eliminate the torsion $T_\a$ from (\ref{N}) by means of
our master equation (\ref{master_1a}). After some algebra we find
\begin{equation}\label{dist}
\!  {N}_{\alpha\beta} = {1\over 2}Q_{\alpha\beta} -2\left(
    {\xi}_{0}-{1\over 2}\right)\qslash_{[ \alpha\beta ] \gamma }
  {\vartheta}^{\gamma} -2\left( {\xi}_{0}+ {\xi}_{1} - {1\over 2}
  \right) Q_{[\alpha} {\vartheta}_{\beta ]} -{1\over 2}e_{[ \alpha}
  \rfloor  {^{(3)}T_{\beta ]}}.
\end{equation}
Note that $e_\a\rfloor Q_{\b\g}=Q_{\a\b\g}$ and $e_\a\rfloor Q=Q_\a$.
Moreover, by means of (\ref{master_1a}), we can also express the first
two irreducible pieces of the torsion (\ref{decomp_torsion}) and
(\ref{tor2}) in terms of the nonmetricity,
\begin{eqnarray}
  ^{(1)}T^{\alpha} & = & {\xi}_{0}( {\qslash}^{\alpha}{}_{\mu}\wedge
  {\vartheta}^{\mu}+\frac{1}{3}{\Lambda}\wedge {\vartheta}^{\alpha}) =
  {\xi}_{0}P^{\alpha}\, ,\label{1T} \\
 ^{(2)}T^{\alpha} & = &
  -\frac{1}{3}\left[ {\xi}_{0}{\Lambda}-3({\xi}_{0}
    +{\xi}_{1})Q\right]\wedge {\vartheta}^{\alpha}\, ,\label{2T}
\end{eqnarray}
with the 2-form $P^{\alpha}$ of (\ref{P_alpha}). Note that both,
$^{(1)}T^{\alpha}$ and $P^\a$, have 16 independent components.
Eq.(\ref{2T}) is equivalent to (\ref{traceansatz}).

Further insight into the structure of the metric-affine field
equations can be gained if we take care of the master equation
(\ref{master_1}) in the excitations (\ref{M-excit}) and
(\ref{Ha-excit}).  Let us first turn to the simpler expression
(\ref{Ha-excit}). With our master equation, we derived $^{(1)}T^\a$
and $^{(2)}T^\a$ in (\ref{1T}) and (\ref{2T}), respectively. We
substitute these two pieces, together with the irreducible
decompositions (\ref{deco4}), (\ref{3Q}), and (\ref{Q2}), into
(\ref{Ha-excit}). We find
\begin{eqnarray}\label{Ha-excit'}
  -\kappa H_\a&=&{}^\star\left(a_1{}^{(1)}T_\a + a_2{}^{(2)}T_\a+
    a_3{}^{(3)}T_\a+ c_2{}^{(2)}Q_{\a\b}\wedge \vta^\b +
    c_3{}^{(3)}Q_{\a\b}\wedge \vta^\b\right.\cr &&\left.\hspace{14pt}
    + c_4{}^{(4)}Q_{\a\b}\wedge \vta^\b \right)\cr &=&
  {}^\star\left\{a_1\xi_0 P_\a-\frac{a_2}{3}\left[\xi_0\Lambda-3
      (\xi_0+\xi_1) Q\right]\wedge\vta_\a+a_3{}^{(3)}T_\a
    -\frac{2c_2}{3}\left(e_{(\a}\rfloor P_{\b)}
    \right)\wedge\vta^\b\right. \cr &&\left.
    \hspace{14pt}+\frac{4c_3}{9}\left( \vta_{(\a}
      e_{\b)}\rfloor\Lambda- \frac{1}{4}g_{\a\b} \Lambda\right)
    \wedge\vta^\b +c_4Q\wedge\vta_\a \right\}\,.
\end{eqnarray}
Now we order the right-hand side in terms of $P_\a$, $\Lambda$, and
$Q$. After some algebra, we have
\begin{eqnarray}\label{Ha-excit''}
  -\kappa H_\a&=&{}^\star\left\{(a_1\xi_0+c_2)P_\a-\frac 13(a_2\xi_0
    +c_3)\Lambda\wedge \vta_\a +\left[a_2(\xi_0+\xi_1)
      +c_4\right]Q\wedge\vta_\a\right.\cr&&\left.\hspace{14pt}+
    a_{3}{}^{(3)}T^\a\right\}\,.
\end{eqnarray}

The evaluation of the excitation (\ref{M-excit}) is a bit more
complicated. In expanded form, eq.(\ref{M-excit}) reads
\begin{eqnarray}\label{M-excit'}
  -\frac{\kappa }{2} M^{\alpha\beta} & = &{} ^{\star }\! \left( b_{1}
    {^{(1)}Q}^{\a\b}+ b_{2} {^{(2)}Q}^{\a\b}+ b_{3} {^{(3)}Q}^{\a\b}+
    b_{4} {^{(4)}Q}^{\a\b} \right) \cr & & +
  c_{2}{\vartheta}^{({\alpha}} \wedge {} ^{\star (1)}T^{\beta )} +
  c_{3}{\vartheta}^{({\alpha}}\wedge {} ^{\star (2)}T^{\beta )} +
  \frac{1}{4}(c_{3}-c_{4})\,^{\star }Tg^{\alpha\beta} \cr & &
  +\frac{b_{5}}{2}\left[ {\vartheta}^{( \alpha}\wedge
    {}^{\star}(Q\wedge {\vartheta}^{\beta )})-\frac{1}{4}
    g^{\alpha\beta}{}\,^{\star}(3Q+{\Lambda})\right]\,.
\end{eqnarray}
Now we substitute into this equation the irreducible pieces
$^{(2)}Q_{\a\b}, {}^{(3)}Q_{\a\b}, {}^{(4)}Q_{\a\b}$ from (\ref{Q2}),
(\ref{3Q}),(\ref{deco4}), respectively, $^{(1)}T^\a, {}^{(2)}T^\a$
from (\ref{1T}), (\ref{2T}), and $T$ from (\ref{traceansatz}):
\begin{eqnarray}\label{M-excit''}
  -\frac{\kappa }{2}
  M^{\alpha\beta}&=&b_1{}^{\star(1)}Q^{\a\b}-\frac{2b_2}{3}
  {}^\star\left( e^{(\a}\rfloor
    P^{\b)}\right)+\frac{4b_3}{9}\left[^\star\left(\vta^{(\a}e^{\b)}
      \rfloor \Lambda\right)-\frac{1}{4}g^{\a\b}{}^\star\Lambda
  \right]\cr&& +b_4{}^\star Q g^{\a\b}+c_2\xi_0\vta^{(\a}
  \wedge{}^\star P^{\b)}-\frac{c_3}{3}\vta^{(\a}\wedge{}^\star\left\{
    \left[\xi_0 \Lambda-3(\xi_0+\xi_1)Q
    \right]\wedge\vta^{\b)}\right\}\cr && +
  \frac{c_3-c_4}{4}\left[\xi_0{}^\star \Lambda-3(\xi_0+\xi_1) {}^\star
    Q \right]g^{\a\b}\cr && +\frac{b_{5}}{2}\left[ {\vartheta}^{(
      \alpha}\wedge( e^{\b)}\rfloor{}^{\star}Q)-\frac{1}{4}
    g^{\alpha\beta}(3{}^{\star}Q +{}^{\star}{\Lambda})\right]\,.
\end{eqnarray}
With some algebra\footnote{Note that for any 1-form
  $\Phi=\Phi_\g\vta^\g$ we have $^\star\left[\vta^\a\left(e^\b\rfloor
      \Phi \right) \right]=\Phi^{\b}{}^\star \vta^{\a}=
  \Phi^\b\eta^\a$. A bit more complicated is the computation of
  $\vta^\a\wedge{}^\star\left( \Phi\wedge\vta^\b \right)$. If $\Omega$
  is another 1-form, we have the general rule $^\star\Phi\wedge
  \Omega={}^\star\Omega\wedge\Phi $. Moreover, we have the rules for
  the Hodge star for any form $^\star\left(\Phi \wedge \vta^\a
  \right)=e^\a \rfloor{} ^\star\Phi$ and (in four dimensions)
  $^\star\left(e^\a\rfloor \Phi \right)=-{}^\star\Phi \wedge \vta^\a
  $. Consequently,
\begin{eqnarray*}\label{M-exci'''}
  \vta^\a\wedge{}^\star\left( \Phi\wedge\vta^\b \right)&
  =&\vta^\a\wedge \left(e^\b\rfloor{}^\star\Phi \right)= - e^\b\rfloor
  \left(\vta^\a \wedge{}^\star\phi \right) + g^{\a\b} {}^\star\Phi =
  e^\b\rfloor \left(^\star\vta^\a \wedge{}\phi \right) + g^{\a\b}
  {}^\star\Phi\cr &= &\eta^{\a\b}\wedge\Phi-\eta^\a\Phi^\b + g^{\a\b}
  {}^\star\Phi\,.
\end{eqnarray*}
Upon symmetrization, we find $\vta^{(\a} \wedge{}^\star\left(
  \Phi\wedge\vta^{\b)} \right)= -\eta^{(\a}\Phi^{\b)} + g^{\a\b}
{}^\star\Phi\,.$} we can order $M^{\a\b}$ in a similar way as we did
with $H_\a$:
\begin{eqnarray}\label{M-excit'''}
  &&\hspace{-15pt} -\frac{\kappa }{2} M^{\alpha\beta}
  =b_1{}^{\star(1)}Q^{\a\b}+\frac 13\left({2b_2}+3c_2\xi_0 \right)
  \vta^{(\a}\wedge{}^\star P^{\b)}\cr && +\frac 19\left(4b_3+3c_3\xi_0
  \right)\Lambda^{(\a} \eta^{\b)}-\frac{1}{72}\left(8b_3+ 18 c_4
    \xi_0+ 6 c_3\xi_0+9b_5\right)g^{\a\b} \Lambda_\mu\eta^\mu\\ && -
  \frac 12\left[2c_3(\xi_0+\xi_1)+b_5 \right]Q^{(\a}\eta^{\b)} +\frac
  18\left[8b_4+2(c_3+3c_4)(\xi_0 + \xi_1)+b_5
  \right]g^{\a\b}Q_\mu\eta^\mu\,.\nonumber
\end{eqnarray}
This completes our simplifications of $H_\a$ and $M^{\a\b}$.

Incidentally, the distortion (\ref{dist}) exhibits the special role of
the choice ${\xi}_{0}=1/2$, ${\xi}_{1}=0$ or $Q=0$, and $^{(3)}
T^{\alpha}=0$.  In either case, the connection reduces to
\begin{equation}\label{eq19}
  {\Gamma}_{\alpha\beta} = \widetilde{\Gamma}_{\alpha\beta} + {1\over
    2}Q_{\alpha\beta}\,, \quad\quad\hbox{with}\quad {\xi}_{0}
  =\frac{1}{2},\> {\xi}_{1}=0\quad \hbox{or}\quad Q=0 \,,\>{}
  ^{(3)}T^{\alpha}=0\,.
\end{equation}
Metric-affine spacetimes with such a simple connection have already
been studied before.\footnote{These spacetimes emerge in the following
  context: We define the Palatini 3-form ${\cal
    P}^\a{}_\b:=-\delta(\eta_{\mu\nu}\wedge
  R^{\mu\nu}/2)/\delta\Gamma_\a{}^\b$ and find ${\cal P}_{\a\b}=-{\cal
    P}_{\b\a}=D\eta_{\a\b}/2=-Q\wedge \eta_{\a\b}+T^\g\wedge
  \eta_{\a\b\g}/2$. If we require $e_{[\a}\rfloor {\cal
    P}^\g{}_{\b]}=0$, then\cite{HehlKerlick_1978}
  ${\Gamma}_{\alpha\beta} = \widetilde{\Gamma}_{\alpha\beta}
  +Q_{\alpha\beta}/2$.} We will come back to such a connection later
in the discussion of our new exact solution in Sec.\ref{displaysol}.

Eventually, we can also substitute the master equation
(\ref{master_1}) and the choice (\ref{exotic}) into the Lagrangian
(\ref{QMA}). Again, like with the excitations $H_\alpha$ and
$M^{\alpha\beta}$, we express the Lagrangian in terms of $P^\alpha$,
$\Lambda$, and $Q$. We find 
\begin{eqnarray}
  V&  = & \frac{1}{2\kappa}\left\{ -a_{0}R^{\alpha\beta}\wedge
    {\eta}_{\alpha\beta} - 2{\lambda}_{0}{\eta} + b_{1}\, ^{\star
      (1)}Q^{\alpha\beta}\wedge\,  ^{(1)}Q_{\alpha\beta} \right. \cr & &
  \cr & & + \left[ (a_{1}{\xi}_{0}+c_{2}){\xi}_{0} +
    \frac{1}{3}(2b_{2}+3c_{2}{\xi}_{0})\right] P^{\alpha}\wedge \,
  ^{\star}P_{\alpha} \cr & & \cr & & + \frac{1}{3}\left[
    (a_{2}{\xi}_{0}+c_{3}){\xi}_{0}+c_{3}{\xi}_{0}+\frac{4}{3}b_{3}\right]
  {\Lambda}^{\mu}{\Lambda}_{\mu}{\eta}\cr & & \cr & & +
  [3(a_{2}{\xi}_{0}+c_{4}){\xi}_{0}+3c_{4}{\xi}_{0}+4b_{4}]Q^{\mu}Q_{\mu}{\eta}\cr
  & & \cr & & - \left. \left[
      2(a_{2}({\xi}_{0}+{\xi}_{1})+c_{4}){\xi}_{0}+2(a_{2}{\xi}_{0}+ 
      c_{3})({\xi}_{0}+{
        \xi}_{1})-\frac{4}{3}a_{2}{\xi}_{0}({\xi}_{0}+{\xi}_{1}) +
      b_{5}\right]\right. \cr & & \cr & &\left.  
\times Q^{\mu}{\Lambda}_{\mu}{\eta}\right\}+
  V_{R^2}\,.
\end{eqnarray}
We put here $a_3=0$.

\section{Finding solutions by nullifying the excitations}

We can find exact solutions of the field equations of MAG
straightforwardly in a very simple manner.  We will ask for
non-trivial field configurations with the property of vanishing field
excitations, i.e., we will require
\begin{equation}\label{momenta_0}
  H_{\alpha}  =  0 \, , \qquad M^{\alpha\beta}  =  0 \, , \qquad
  H^{\alpha}{}_{\beta}  =  0 \, .
\end{equation}
And indeed, because of the inhomogeneity of the excitations in terms
of the field strengths, it will be possible to generate solutions with
non trivial curvature.\bigskip

\centerline{{\it Table 2.} The case $Q=0$.}\vspace{-5pt}
$$
{\begin{tabular}{|c||c|} \hline\label{Table_2}
Excitation & constraints\\
\hline\hline
   & $a_{1}{\xi}_{0}+c_{2}=0$\\
    $H_{\alpha}=0$     & $a_{2}{\xi}_{0}+c_{3}=0$ \\
         & $a_{3}=0 \quad {\rm or}\quad ^{(3)}T^{\alpha}=0$  \\
\hline\hline
  & $b_{1}=0$\\
 $M^{\alpha\beta}=0$ & $2b_{2}+3c_{2}{\xi}_{0}=0$\\
  & $4b_{3}+3c_{3}{\xi}_{0}=0$\\
  & $b_{5}+2c_{4}{\xi}_{0}=0$\\
\hline\hline
       & $w_{1}=w_{2}=w_{3}=w_{4}=w_{5}=0$\\
  $H^{\alpha}{}_{\beta}=0$  & $z_{1}=z_{2}=z_{3}=z_{4}=z_{5}=0$\\
       & $6\rho a_{0}-{\kappa}w_{6}W=0$ \\
\hline \hline
\end{tabular}}
$$
\medskip

If we substitute this into the sourcefree field equations
(\ref{first}) and (\ref{second}), only the following truncated
equation is left over:
\begin{eqnarray}\label{algebraic}
 E_{\alpha} & = & e_{\alpha}\rfloor V = 0\label{E_a}\,.
\end{eqnarray}
Since $\vartheta_{[\alpha}\wedge E_{\beta ]}=0$, this equation has
only 10 independent components. The field equation (\ref{zeroth}) is
redundant because (\ref{first}) and (\ref{second}) are fulfilled.
Accordingly, we have just to solve the algebraic relation
(\ref{algebraic}).

Let us now have a look at the excitations given in equations
(\ref{Ha-excit''}), (\ref{M-excit'''}), and (\ref{H_ab}),
respectively, and ask under which conditions these excitations will
vanish without being dynamically trivial. Naturally, we can
distinguish between the different cases $Q=0$ and $Q\neq 0$. In the
case of vanishing Weyl covector $Q$, the conditions are collected in
Table 2, for nonvanishing $Q$ in Table 3.

Having the conditions at our disposal that are listed in the Tables 2
and 3, the construction of exact solutions of MAG is appreciably
simplified.  The {\em only} equation that has to be fulfilled is
eq.(\ref{E_a}).
\bigskip

\centerline{{\it Table 3.} The case $Q\neq 0$.}
$$
{\begin{tabular}{|c||c|} \hline\label{Table_3}
Excitation & constraints\\
\hline\hline
   & $a_{1}{\xi}_{0}+c_{2}=0$\\
    $H_{\alpha}=0$     & $a_{2}{\xi}_{0}+c_{3}=0$ \\
     & $a_{2}({\xi}_{0}+{\xi}_{1})+c_{4}=0$\\
         & $a_{3}=0 \quad {\rm or}\quad ^{(3)}T^{\alpha}=0$  \\
\hline\hline
  & $b_{1}=0$\\
  & $2b_{2}+3c_{2}{\xi}_{0}=0$\\
  $M^{\alpha\beta}=0$ & $4b_{3}+3c_{3}{\xi}_{0}=0$\\
  & $4b_{4}+3c_{4}({\xi}_{0}+{\xi}_{1})=0$\\
  & $b_{5}+2c_{3}({\xi}_{0}+{\xi}_{1})=0$\\
  & $c_{4}{\xi}_{0}-c_{3}({\xi}_{0}+{\xi}_{1})=0$\\
\hline\hline
       & $w_{1}=w_{2}=w_{3}=w_{4}=w_{5}=0$\\
  $H^{\alpha}{}_{\beta}=0$  & $z_{1}=z_{2}=z_{3}=z_{4}=z_{5}=0$\\
       & $6\rho a_{0}-{\kappa}w_{6}W=0$ \\
\hline \hline
\end{tabular}}
$$
\medskip

In this article, we will concentrate on Table 2, that is, on the case
of {\it vanishing Weyl covector}, $Q=0$. In particular, we have
$b_1=0$ and we choose the option $a_3=0$. Then the constraints of
Table 2 can be used to eliminate the constants $a_1,a_2,a_3$ and
$b_2,b_3,b_5$ from the gauge Lagrangian (\ref{QMA}) obeying the
conditions (\ref{exotic}):
\begin{eqnarray}\label{VI}
  V & = & \frac{1}{2\kappa}\left[
    -a_0 R^{\alpha\beta}\wedge{\eta}_{\alpha\beta}-2{\lambda}_{0}{\eta}
  \right. \nonumber\\ & & \left.\quad\quad -\frac{1}{\xi_0}\, T^{\alpha}\wedge{}
    ^{\star}\left( c_{2}{} ^{(1)}T_{\alpha}+c_{3}{}
      ^{(2)}T_{\alpha}\right)\right.  \nonumber\\ & &\left. \quad\quad
    + 2\left( c_{2}{}^{(2)}Q_{\alpha\beta}
      + c_{3}{}^{(3)}Q_{\alpha\beta}\right)\wedge {\vartheta}^{\alpha}
    \wedge {}^{\star}T^{\beta}\right.  \nonumber\\& & \left.\quad\quad
    -\frac{3\xi_0}{4}Q_{\alpha\beta}\wedge {}^{\star}\left(
      2c_{2}{}^{(2)}Q^{\alpha\beta}
      +c_{3}{}^{(3)}Q^{\alpha\beta}\right) \right] \nonumber\\ & & -
  \frac{w_{6}}{2\rho}\,R^{\alpha\beta}\wedge {} ^{\star (6)}W_{\alpha\beta}\,.
\end{eqnarray}
We also dropped $Q$-dependent terms. This is equivalent to $c_4=0$
and, according to Table 2, corresponds to $b_5=0$.

We still didn't apply the last constraint of Table 2. Now we can use
it in order to eliminate the constant $w_6$. But we first collect the
curvature dependent terms in (\ref{VI}). For this purpose we recall
the geometric identities
\begin{equation}
  R^{\alpha\beta}\wedge {\eta}_{\alpha\beta}=\,
  ^{(6)}W^{\alpha\beta}\wedge 
  {\eta}_{\alpha\beta}=-W{\eta}\quad {\rm
    and}\quad R^{\alpha\beta}\wedge \, ^{\star (6)}W_{\alpha\beta}=
  \frac{W^2}{12}{\eta}\,. 
\end{equation}
Then, we have quite generally
\begin{equation}\label{V_R}
  -\frac{a_{0}}{2\kappa}\,R^{\alpha\beta}\wedge{\eta}_{\alpha\beta} 
  -\frac{w_6}{2\rho}\, 
  R^{\alpha\beta}\wedge{} ^{\star(6)}W_{\alpha\beta}=\frac{1}{2\kappa}\, W
  \left(a_0-\frac{\kappa w_6}{12\rho}W\right)\eta\, ,
\end{equation}
and the Lagrangian (\ref{VI}) can be rewritten as
\begin{eqnarray}\label{VI'}
  V & = & \frac{1}{2\kappa}\left[ W
  \left(a_0-\frac{\kappa w_6}{12\rho}\,W\right)\eta\,
   -2{\lambda}_{0}{\eta}
  \right. \nonumber\\ & & \left.\quad\quad -\frac{1}{\xi_0}\, T^{\alpha}\wedge{}
    ^{\star}\left( c_{2}{} ^{(1)}T_{\alpha}+c_{3}{}
      ^{(2)}T_{\alpha}\right)\right.  \nonumber\\ & &\left. \quad\quad
    + 2\left( c_{2}{}^{(2)}Q_{\alpha\beta}
      + c_{3}{}^{(3)}Q_{\alpha\beta}\right)\wedge {\vartheta}^{\alpha}
    \wedge {}^{\star}T^{\beta}\right.  \nonumber\\& & \left.\quad\quad
    -\frac{3\xi_0}{4}Q_{\alpha\beta}\wedge {}^{\star}\left(
      2c_{2}{}^{(2)}Q^{\alpha\beta}
      +c_{3}{}^{(3)}Q^{\alpha\beta}\right) \right]
\,.
\end{eqnarray}
We recognize that in the curvature dependent pieces we have a weak
gravity contribution $\sim W/\kappa$ and a strong gravity contribution
$\sim W^2/\rho$. This is the final Lagrangian for the field equations
of which we find an exact solutions. However, we still have to fulfill
the last constraint of Table 2, namely
\begin{equation}\label{Wconst}
\kappa w_6W=6\rho a_0\,. 
\end{equation}
This means that the curvature scalar $W$ of our solution is required
to be a {\it constant.} Substituting the constraint into (\ref{VI'}),
the expression in the first parentheses becomes $a_0/2$, i.e., only a
linear and constant term in $W$ is left over, namely
$(a_0W/4\kappa)\,\eta$.

We could find a relatively trivial solution by putting $a_0=0$ and
$\lambda_0=0$, but that is not our desire, see, however, Adak and
Sert\cite{Adak2004}. Therefore, we rather choose $a_0=1$ in future.
Then our Lagrangian (\ref{VI'}) depends, besides the (weak)
gravitational constant $\kappa$ and the cosmological constant
$\lambda_0$, only on the arbitrary parameters $\xi_0$ and $c_2,c_3$.

\section{Seed Solution carrying metric and torsion}

The formulation of Einstein's field equation in terms of differential
ideals\cite{Guerses_a} opens the possibility to create non-trivial
solutions of the field equation even by starting from flat Minkowski
spacetime. The Kerr-Schild transformation of general relativity
provides another example. The method of prolongations suggests, as we
have seen, an ansatz in form of the master equation (\ref{master_1}).
We choose as a seed solution a suitable coframe ${\vartheta}^{\alpha}$
(or metric $g_{\alpha\beta}$) and a torsion $T^{\alpha}$ and impose
further assumptions or constraints that will lead to purely algebraic
equations, which will be nonlinear in general. But remember, we could
start with {\em any} metric and {\em any} torsion as a seed solution
and even in the presence of matter this methods works. We don't need
to take recourse to the field equations that the seed solutions have
to obey, the field equations of MAG alone will determine the relevant
geometrical quantities.

Of course, we will start from seed solutions that are expected to be
of physical relevance, such as the Kerr metric given in
Ref.\cite{Chandra_82} and the torsion displayed in Ref.\cite{BaeKerr}.
In the case of vanishing nonmetricity $Q^{\alpha\beta}$, our solution
to be found should go over into a solution of the Poincar\'e gauge field theory
(PGT) and further limits to be taken will lead to Newton-Einstein
gravity, provided the coupling constants will be adjusted suitably.

Accordingly, for our purposes, we choose a two-step procedure: We
first take a metric $g$ of an exact solution of general relativity and
then, keeping the metric fixed, turn on the torsion $T^\a$ by going
over to a known solution of the field equations of the Poincar\'e
gauge theory, in which, as we recall, the nonmetricity vanishes
identically.\footnote{Presentations of the Poincar\'e gauge theory can
  be found in Nester\cite{NesterDr}, Blagojevi\'c\cite{Milutin}, and
  Gronwald et al.\cite{Erice95}, for recent results one should compare
  Obukhov\cite{gaugegr} and, for possible observations, Preuss et
  al.\cite{Preuss2005}; the Kerr solution and its approximation in a
  telelparallel spacetime was discussed by Pereira, Vargas, and
  Zhang\cite{Pereira2001,Zhang2005}.} Taking this as a new starting
point, we eventually switch on the nonmetricity $Q_{\a\b}$ and
generate a whole class of solutions of MAG.

Let us start with the {\it Kerr-deSitter metric\/} of general
relativity with cosmological constant ${\lambda}$ and an exact
solution of the Poincar\'e gauge theory using that metric and
providing additionally the torsion by solving the field
equations of the Poincar\'e gauge theory.\cite{BaeKerr}

\subsection{Seed metric as solution of Einstein's field equation}

The coframe ${\vartheta}^{\alpha}$, in terms of coordinates
$t,r,\theta,\phi$, reads
\begin{eqnarray}
{\vartheta}^{0} & = & \sqrt{\frac{{\Delta}}{\Sigma}}(dt +
a{\sin^{2} {\theta}} d {\phi} )\,,\cr
 {\vartheta}^{1} & = &
\sqrt{\frac{{\Sigma}}{\Delta}}dr \,,\cr
 {\vartheta}^{2} & = &
\sqrt{\frac{{\Sigma}}{F}}d{\theta}\,,\cr
 {\vartheta}^{3} & =
& {\sin \theta}\sqrt{ \frac{F}{\Sigma}}\left[ adt +
(r^2+a^2)d{\phi}\right]\,. \label{COFRAME}
\end{eqnarray}
The structure functions are defined according to
\begin{eqnarray}\label{DSF}
{\Sigma} & := & r^2 + a^2{\cos^{2} \theta}\, ,\\
F & := & 1 + \frac{1}{3}{\lambda}a^2{\cos^2 \theta}\, , \\
{\Delta} & := & r^2+a^2-2Mr-\frac{1}{3}{\lambda}r^2(r^2+a^2)\,
.\label{DSF'}
\end{eqnarray}
The coframe is orthonormal. Then the metric reads
\begin{equation}\label{anhmetric}
  g = -{\vartheta}^{0}\otimes {\vartheta}^{0} + {\vartheta}^{1}\otimes
  {\vartheta}^{1} + {\vartheta}^{2}\otimes {\vartheta}^{2} +
  {\vartheta}^{3}\otimes {\vartheta}^{3}
\end{equation}
or, in terms of local coordinates,
\begin{eqnarray}
  g = -\frac{\Delta}{\Sigma}\left( dt +
    a{\sin^{2}\theta}d{\phi}\right)^2 +
  \frac{\Sigma}{\Delta}dr^2+\frac{\Sigma}{F}d{\theta}^2
  {\sin^2\theta}+\frac{F}{\Sigma}\left[ adt +
    (r^2+a^2)d{\phi}\right]^{2}.
\end{eqnarray}
This solution of Einstein's field equation depends on the set of
essential constants $\{M, a, {\lambda}\}$, i.e., on {\it mass, angular
  momentum\/} per mass, and the {\it cosmological constant.}

In the limit of vanishing Kerr-parameter $a\rightarrow 0$, the
coframe (\ref{COFRAME}) reduces to the Schwarzschild-deSitter (also
known as Kottler) coframe with
\begin{eqnarray}
{\vartheta}^{0} & = & \left(
1-\frac{2M}{r}-\frac{\lambda}{3}r^2\right)^{1/2}dt\, , \cr
{\vartheta}^{1} & = & \left(
1-\frac{2M}{r}-\frac{\lambda}{3}r^2\right)^{-1/2}dr\, , 
\cr {\vartheta}^{2} & = & rd{\theta}\, , \cr {\vartheta}^{3} & = &
r{\sin\theta}d{\phi}\, ,
\end{eqnarray}
whereas in the limit of a vanishing cosmological constant
${\lambda}\rightarrow 0$ we recover the well-known
Kerr-solution.\cite{Kerr_1963a} Further physical and mathematical
properties of the solution are compiled in the book of
Chandrasekhar.\cite{Chandra_82}

The Riemannian curvature ${\tilde R}^{\alpha\beta}$ comprises three
irreducible pieces, the Weyl-curvature, the tracefree symmetric Ricci
and the curvature scalar:
\begin{equation}
{\tilde R}^{\alpha\beta} = {}^{(1)}{\widetilde W}^{\alpha\beta}
+ {}^{ (4)}{\widetilde W}^{\alpha\beta} + {}^{
(6)}{\widetilde W}^{\alpha\beta}\,.
\end{equation}
The numbers $(1),(4),(6)$ refer already to a metric-affine space in
which the rotational curvature $W_{\a\b}:=R_{[\a\b]}$ has six
independent components, see (\ref{wewe}) to (\ref{w1w1}). In a
Riemannian space only the pieces with the numbers $(1),(4),(6)$ are
non-vanishing. Explicitly, these matrices are given by
\begin{eqnarray}
^{(1)}{\widetilde W}^{\alpha\beta} & = &
\frac{Mr}{{\Sigma}^3}(r^2-3a^2{\cos^2\theta}) \pmatrix{0 &
-2{\vartheta}^{01} & {\vartheta}^{02} & {\vartheta}^{03} \cr
         \diamond & 0 & {\vartheta}^{12} & {\vartheta}^{13} \cr
         \diamond & \diamond & 0 & -2{\vartheta}^{23} \cr
         \diamond & \diamond & \diamond & 0}\cr & & \cr & &
+\frac{Ma\,{\cos\theta}}{{\Sigma}^3}(3r^2-a^2{\cos^2\theta})
\pmatrix{0 & -2{\vartheta}^{23} & -{\vartheta}^{13} &
{\vartheta}^{12} \cr \diamond & 0 & -{\vartheta}^{03} &
{\vartheta}^{02} \cr \diamond & \diamond & 0 & 2{\vartheta}^{01}
\cr \diamond & \diamond & \diamond & 0  }\,,\\
^{(4)}{\widetilde W}^{\alpha\beta} & = & 0\,, \label{ricf} \\
^{(6)}{\widetilde W}^{\alpha\beta} & = &
-\frac{\lambda}{3}{\vartheta}^{\alpha\beta}\,. \label{WEYL-V4}
\end{eqnarray}
The symbols $\diamond$ and $\bullet$ denote matrix elements already
known because of the antisymmetry or the symmetry of the matrix
involved. The matrices (\ref{ricf}) and (\ref{WEYL-V4}) are equivalent
to the statement that Einstein's vacuum field equation with
cosmological constant ${\lambda}$ is fulfilled for the metric
(\ref{anhmetric}) and the coframe (\ref{COFRAME}).

A canonical form of the metric for the most general type D solution of
the Einstein-Maxwell equations (with cosmological constant) has been
given by Pleba\'n\-ski and Demia\'nski\cite{Plebanski_1976} resulting
in a {\it seven\/} parameter solution. This was updated by Debever,
Kamran, and McLenaghan,\cite{Debever83,Debever84} see also
Grac\'{\i}a and Mac\'{\i}as.\cite{GarciaGR} This solution could be
taken as seed solution as well. However, because of simplicity, we
will concentrate on the metric as given in (\ref{anhmetric}), together
with the coframe (\ref{COFRAME}) and the structure functions
(\ref{DSF}) to (\ref{DSF'}).

\subsection{Seed torsion of the Poincar\'e gauge theory}

The torsion $T^{\alpha}$ of a stationery axially symmetric solution of
the Poincar\'e gauge theory reads\cite{BaeKerr}
\begin{eqnarray}
  T^{0} & = & \sqrt{\frac{{\Sigma}}{\Delta}}\left[
    -v_{1}{\vartheta}^{01} +
    \sqrt{\frac{{\Sigma}}{\Delta}}\left[v_{2}({\vartheta}^{02}
      -{\vartheta}^{12})+ v_{3}({\vartheta}^{03}-{\vartheta}^{13})\right]
      -2v_{4}{\vartheta}^{23}\right]\,, \cr
    T^{1} & = & T^{0}\,,
    \cr T^{2} & = & \sqrt{\frac{{\Sigma}}{\Delta}}\left[
      v_{5}({\vartheta}^{02}-{\vartheta}^{12})+
      v_{4}({\vartheta}^{03}-{\vartheta}^{13})\right]\,,\cr 
    T^{3} & = & \sqrt{\frac{{\Sigma}}{\Delta}}\left[
      -v_{4}({\vartheta}^{02}-{\vartheta}^{12})+
      v_{5}({\vartheta}^{03}-{\vartheta}^{13})\right]\,,
\label{TORSION-U4}
\end{eqnarray}
together with the functions
\begin{eqnarray}
 v_{1} & = & \frac{M}{{\Sigma}^2}(r^2-a^2{\cos^{2}\theta})\,,\cr 
 v_{2} & = & -\frac{Ma^{2}r{\sin\theta}{\cos\theta}}{{\Sigma}^2}
 \sqrt{\frac{F}{\Sigma}}\,,\cr
 v_{3} & = &
 \frac{Mar^2{\sin\theta}}{{\Sigma}^{2}}\sqrt{\frac{F}{\Sigma}}\,,\cr
 v_{4} & = & \frac{Mar\,{\cos\theta}}{{\Sigma}^2}\,,\cr
 v_{5} & = & \frac{Mr^2}{{\Sigma}^2} \label{torsionsubs}
\end{eqnarray}
and the coframe (\ref{COFRAME}). Note that also the choice
$T^{1}=-T^{0}$ would lead to a viable solution of MAG.

For the torsion trace $T= e_{\alpha}\rfloor T^{\alpha}$, we find
\begin{equation}
  T  = \sqrt{ \frac{{\Sigma}}
    {\Delta}}(v_{1}-2v_{5})({\vartheta}^{0}-{\vartheta}^{1})\,,
\end{equation}
and the axial torsion, see (\ref{tor3}), turns out to vanish:
\begin{equation}\label{axitor=0}
  {^{(3)}T^{\alpha}} = 0\,.
\end{equation}
This completes our seed solution of the Poincar\'e gauge theory. We
now turn to our ansatz for finding a solution of MAG.

\section{Ansatz for the nonmetricity}

The nonmetricity 1-form can be decomposed into components
according to
\begin{equation}\label{nonmetricity1}
Q^{\alpha\beta}=Q_\gamma{}^{\alpha\beta}\vartheta^\gamma\,.
\end{equation}
We can represent it as a symmetric $4\times 4$-matrix with 10
independent components, each of which is a sum of suitable 1-forms:
\begin{equation}\label{nonmetricity2}
  Q^{\alpha\beta}=\pmatrix{ Q^{00}&Q^{01}&Q^{02}&Q^{03}\cr
    \bullet&Q^{11}&Q^{12}&Q^{13}\cr \bullet&\bullet&Q^{22}&Q^{23}\cr
    \bullet&\bullet&\bullet&Q^{33}} =\pmatrix{
    Q_\gamma{}^{00}\vartheta^\gamma&Q_\gamma{} ^{01}
    \vartheta^\gamma&Q_\gamma{}^{02} \vartheta^\gamma&Q_\gamma{}^{03}
    \vartheta^\gamma\cr \bullet&Q_\gamma{}^{11}
    \vartheta^\gamma&Q_\gamma{}^{12} \vartheta^\gamma&Q_\gamma{} ^{13}
    \vartheta^\gamma\cr \bullet&\bullet&Q_\gamma{}^{22}
    \vartheta^\gamma&Q_\gamma{} ^{23} \vartheta^\gamma\cr
    \bullet&\bullet&\bullet&Q_\gamma{}^{33} \vartheta^\gamma}.
\end{equation}
The bullets denote those matrix elements that, due to symmetry, can be
read off from the other matrix elements.

Since we are looking for a stationary axially symmetric solution, it
would appear natural to start with the most general axially symmetric
form of $Q_{\a\b}$. So far, this form is unknown. Moreover, we expect
that it is so general that it would not help us in our task of solving
the field equations of MAG. Therefore, we start from the most general
spherically symmetric form of $Q_{\a\b}$ and generalize it. For
vanishing Kerr parameter $a$, our solution has to reduce to the
spherically symmetric form.

For $SO(3)$-symmetry, by solving the Killing equations,
Tresguerres\cite{Romu1,Romu2} derived the most general form of the
nonmetricity; this has been confirmed by different
groups.\cite{Minkevich95,Ho97,MinkevichVas2003,MAGIII} Tresguerres
found 12 independent components only depending on the radial
coordinate $r$, namely
\begin{eqnarray}\label{nonmetricity5}
  && \hspace{80pt} Q^{\alpha\beta}|_{SO(3)}=\\&&\cr&& \hspace{-10pt}\pmatrix{ {
      Q}_{0}{}^{00} {\vartheta{}^{0}} + {
      Q}_{1}{}^{00}{\vartheta{}^{1}} & { Q}_{0}{}^{01}
    {\vartheta{}^{0}} + { Q}_{1}{}^{01}{\vartheta{}^{1}} & {
      Q}_{2}{}^{02} {\vartheta{}^{2}} + { Q}_{3}{}^{02}
    {\vartheta{}^{3}} &-Q_3{}^{02} {\vartheta{}^{2}} +Q_2{}^{02}
    {\vartheta{}^{3}} \cr \bullet & { Q}_{0}{}^{11} {\vartheta{}^{0}} +
    { Q}_{1}{}^{11} {\vartheta{}^{1}} & { Q}_{2}{}^{12}
    {\vartheta{}^{2}} + { Q}_{3}{}^{12} {\vartheta{}^{3}} &
    -Q_3{}^{12} {\vartheta{}^{2}} + Q_2{}^{12} {\vartheta{}^{3}} \cr
    \bullet &\bullet & { Q}_{0}{}^{22} {\vartheta{}^{0}} + { Q}_{1}{}^{22}
    {\vartheta{}^{1}} & 0 \cr \bullet &\bullet &\bullet &\hspace{8pt}Q_0{}^{22}
    {\vartheta{}^{0}} +Q_1{}^{22} {\vartheta{}^{1}} }.\nonumber
\end{eqnarray}
For convenience, we would like to abbreviate these 12 functions in a
different manner. We translate the holonomic version of Minkevich and
Vasilevski\cite{Minkevich95,MinkevichVas2003} into a corresponding {\em
  an\/}holonomic version. Then, with their ${\frak Q}_0,{\frak
  Q}_1,\dots,{\frak Q}_{11} $, we have
\begin{eqnarray}\label{nonmetricity6}
  Q^{\alpha\beta}|_{SO(3)}\! &=& \!\pmatrix{ {\frak Q}_{0} {\vartheta{}^{0}} +
    {\frak Q}_{1}{\vartheta{}^{1}} & {\frak Q}_{2} {\vartheta{}^{0}} +
    {\frak Q}_{3}{\vartheta{}^{1}} & {\frak Q}_{6} {\vartheta{}^{2}} +
    {\frak Q}_{10}{\vartheta{}^{3}} & - {\frak
      Q}_{10}{\vartheta{}^{2}} + {\frak Q}_{6}{\vartheta{}^{3}} \cr
    \bullet & {\frak Q}_{4} {\vartheta{}^{0}} + {\frak
      Q}_{5}{\vartheta{}^{1}} & {\frak Q}_{7} {\vartheta{}^{2}} +
    {\frak Q}_{11}{\vartheta{}^{3}} & - {\frak
      Q}_{11}{\vartheta{}^{2}} + {\frak Q}_{7}{\vartheta{}^{3}} \cr
    \bullet &\bullet & {\frak Q}_{8} {\vartheta{}^{0}} + {\frak
      Q}_{9}{\vartheta{}^{1}} & 0 \cr \bullet &\bullet &\bullet &
    \hspace{8pt}{\frak Q}_{8} {\vartheta{}^{0}} + {\frak Q}_{9}
    {\vartheta{}^{1}} }\!.\cr &&\cr &&
\end{eqnarray}
In the static $SO(3)$-case, the $\frak{Q}$'s depend only on the radial
coordinate $r$.

Let us now ``blow up'' $Q_{\a\b}$ following considerations as given by
Minkevich and Vasilevski.\cite{MinkevichVas2003} First of all, the
$\frak{Q}$ are assumed to depend on the two variables $r$ and
$\theta$, that is, $\frak{Q}=\frak{Q}(r,\theta)$. Moreover, we need a
few more independent components. For the $SO(3)$-case we have 12
functions. Since the torsion, if its axial piece vanishes, carries 20
independent functions, we tentatively introduce 8 more functions for
the nonmetricity since, according to our master equation
(\ref{master_1}), nonmetricity and torsion are algebraically related.
The following representation of the nonmetricity is a minimal set that
fills our bill,
\begin{eqnarray}
  Q^{\alpha\beta} & = &\! \pmatrix{
    \frak{Q}_{0}{\vartheta}^{0}+\frak{Q}_{1}{\vartheta}^{1} &
    \frak{Q}_{2}{\vartheta}^{0}+\frak{Q}_{3}{\vartheta}^{1} &
    Q^{02} & Q^{03} \cr \bullet &
    \frak{Q}_{4}{\vartheta}^{0}+\frak{Q}_{5}{\vartheta}^{1} &
    Q^{12} & Q^{13} \cr \bullet &\bullet &
    \frak{Q}_{8}{\vartheta}^{0}+\frak{Q}_{9}{\vartheta}^{1} & 0\cr
    \bullet & \bullet & \bullet &
    \frak{Q}_{8}{\vartheta}^{0}+\frak{Q}_{9}{\vartheta}^{1} }\!,
\label{Q_functions}
\end{eqnarray}
with the 1-forms ($\a_1,\a_2,\a_3,\a_4$ are the names in our computer
programs)
\begin{eqnarray}\label{1-forms}
  Q^{02} & = & \underbrace{
    \frak{Y}_{1}{\vartheta}^{0}+\frak{Z}_{1}{\vartheta}^{1}}_{\a_1:=}
  + \frak{Q}_{6}{\vartheta}^{2}+\frak{Q}_{10}{\vartheta}^{3}\,,\cr
  Q^{03} & = & \underbrace{\frak{Y}_{2}{\vartheta}^{0}+
    \frak{Z}_{2}{\vartheta}^{1}}_{\a_2:=}-
  \frak{Q}_{10}{\vartheta}^{2}+\frak{Q}_{6}{\vartheta}^{3}\,,\cr
  Q^{12} & = & \underbrace{\frak{Y}_{3}{\vartheta}^{0}+\frak{Z}_{3}
    {\vartheta}^{1}}_{\a_3:=}+
  \frak{Q}_{7}{\vartheta}^{2}+\frak{Q}_{11}{\vartheta}^{3} \,,\cr
  Q^{13} & = & \underbrace{\frak{Y}_{4}{\vartheta}^{0}+
    \frak{Z}_{4}{\vartheta}^{1}}_{\a_4:=}-
  \frak{Q}_{11}{\vartheta}^{2}+\frak{Q}_{7}{\vartheta}^{3}\,.
\end{eqnarray}
We introduced here 4+4 new functions $\frak{Y}$ and $\frak{Z}$ (or the
four 1-forms $\alpha_1,\cdots,\a_4$), ending up with 20 independent
functions, exactly as planned.  The trace of (\ref{Q_functions})
yields
\begin{equation}
  g_{\a\b}Q^{\alpha\b} = -\frak{q}_1{\vartheta}^{0}
  -\frak{q}_0{\vartheta}^{1}=4Q\,,
\end{equation}
with the abbreviations
\begin{eqnarray}\label{abbr}
  \frak{q}_{0} & := & \frac{1}{4}(\frak{Q}_{1}-\frak{Q}_{5}-
  2\frak{Q}_{9})\,,\qquad \frak{q}_{1}  := \frac{1}{4}(
  \frak{Q}_{0}-\frak{Q}_{4}-2\frak{Q}_{8})\,.
\end{eqnarray}
Hence the expression $Q\wedge \vartheta^\a$, which enters
(\ref{master_1}), reads
\begin{equation}
Q\wedge {\vartheta}^{\alpha}= \frak{q}_{0} \pmatrix{
{\vartheta}^{01}\cr 0\cr -{\vartheta}^{12}\cr -{\vartheta}^{13}}-
\frak{q}_{1} \pmatrix{0\cr {\vartheta}^{01}\cr
{\vartheta}^{02}\cr {\vartheta}^{03} }\,.
\end{equation}

This parameterization of the nonmetricity supports all four irreducible
pieces (\ref{deco4}) to (\ref{deco1}) of $Q_{\a\b}$. Especially, the
two trace-free symmetric second rank tensor pieces $\{
^{(1)}Q^{\alpha\beta}\,,{}^{(2)}Q^{\alpha\beta}\}$ are supported, as
well as the two vector pieces proportional to $\{Q\, , {\Lambda}\}$.
Nevertheless, we will put in (\ref{2cons}), in accordance with Table 2,
the Weyl-covector $Q$ to zero.

\section{Master equation and solutions}


The powerful tool of prolongation theory as applied to the highly
nonlinear partial differential equations of metric-affine gravity
(MAG) results in a set of {\em linear} algebraic equations which
interrelates torsion and nonmetricity. This is our master equation
(\ref{master_1}).

\subsection{Generating nonmetricity}\label{generating}

To generate nonmetricity, we liberate at first the parameter $M$
occurring in (\ref{torsionsubs}), i.e., we let
\begin{equation}\label{relax}
M\longrightarrow L_{0}\quad{\rm in\; eq.(\ref{torsionsubs})}
\end{equation}
in order to introduce a new $GL(4,R)$-charge. Everything else in
(\ref{torsionsubs}) and (\ref{TORSION-U4}) remains untouched. In
particular, the structure functions (\ref{DSF}) to (\ref{DSF'}) keep
their old values. This transformation decouples metric-compatible
(Riemann-Cartan) quantities from metric-affine quantities. Note that
further parameter transformations will lead analogously to viable
nonmetricity functions.

Our tool for generating nonmetricity is the master equation
(\ref{master_1}). We consider its left-hand side specified by the seed
torsion (\ref{TORSION-U4}) and (\ref{torsionsubs}). In particular,
this implies that $^{(3)}T^\a=0$. Then the right-hand side encompasses
the generated nonmetricity. However, we only allow such a nonmetricity
to emerge that is compatible with our ansatz (\ref{Q_functions}) with
(\ref{1-forms}). Only in this way we find equations determining
$Q^{\a\b}$ that remain managable. Altogether, we have then five
functions $v$ for $T^\a$, three functions $\Sigma,F,\Lambda$ for
$\vta^\a$, and 20 functions $\frak{Q,Y,Z}$ for $Q^{\alpha\beta}$.

We substitute (\ref{Q_functions}) and (\ref{1-forms}) into
(\ref{master_1}) and find
\begin{eqnarray}\label{zwtorsion}
  T^{0} & = & {\xi}_{0}\left[
    (\frak{Q}_{1}+\frak{Q}_{2}){\vartheta}^{01}
    +\frak{Y}_{1}{\vartheta}^{02} + \frak{Y}_{2}{\vartheta}^{03} +
    \frak{Z}_{1}{\vartheta}^{12} +
    \frak{Z}_{2}{\vartheta}^{13}-2\frak{Q}_{10}{\vartheta}^{23}\right]
  +{{\xi}_{1}}\frak{q}_{0}{\vartheta}^{01}\,,\cr & & \cr
  T^{1} & = & {\xi}_{0}\left[
    (\frak{Q}_{3}+\frak{Q}_{4}){\vartheta}^{01}
    +\frak{Y}_{3}{\vartheta}^{02} + \frak{Y}_{4}{\vartheta}^{03} +
    \frak{Z}_{3}{\vartheta}^{12} +
    \frak{Z}_{4}{\vartheta}^{13}-2\frak{Q}_{11}{\vartheta}^{23}\right]
  -{{\xi}_{1}}\frak{q}_{1}{\vartheta}^{01}\,,\cr & & \cr
  T^{2} & = & {\xi}_{0}\left[(\frak{Y}_3+\frak{Z}_1)\vta^{01}+
    (\frak{Q}_{6}+\frak{Q}_{8}){\vartheta}^{02}
    +\frak{Q}_{10}{\vartheta}^{03} + (\frak{Q}_{9}-
    \frak{Q}_{7}){\vartheta}^{12}-
    \frak{Q}_{11}{\vartheta}^{13}\right]\cr &&-{{\xi}_{1}}\left(
    \frak{q}_1{\vartheta}^{02} +
    \frak{q}_{0}{\vartheta}^{12}\right)\,,\cr & & \cr
T^{3} & = &
  {\xi}_{0}\left[(\frak{Y}_4+\frak{Z}_2)\vta^{01}
    -\frak{Q}_{10}{\vartheta}^{02}
    +(\frak{Q}_{6}+\frak{Q}_{8}){\vartheta}^{03}
    +\frak{Q}_{11}{\vartheta}^{12} +
    (\frak{Q}_{9}-\frak{Q}_{7}){\vartheta}^{13}\right]\cr
  &&-{{\xi}_{1}}\left( \frak{q}_{1}{\vartheta}^{03} +
    \frak{q}_{0}{\vartheta}^{13}\right)\,.
\end{eqnarray}
If we substitute the torsion (\ref{TORSION-U4}) into
(\ref{zwtorsion}), this represents an under determined system of {\it
  linear\/} algebraic equations for determining the unknown functions
$\frak{Q,Y,Z}$. By comparing the coefficients of the 2-forms
$\vta^{01},\vta^{02},\vta^{03},\vta^{12},\vta^{13},\vta^{23}$, we find
for $T^0$ the 6 equations
\begin{eqnarray} \label{Q-T-1}
  {\xi}_{0}(\frak{Q}_{1}+\frak{Q}_{2}) & = & -\sqrt{\frac{\Sigma}
    {\Delta}}v_{1} -{\xi}_{1}\frak{q}_{0}\,,\qquad
  {\xi}_{0}\frak{Y}_{1} = \frac{\Sigma}{\Delta}v_{2}\,,\qquad
{\xi}_{0}\frak{Y}_{2}  =  \frac{\Sigma}{\Delta}v_{3}\,, \cr
{\xi}_{0}\frak{Z}_{1} & = & -\frac{\Sigma}{\Delta}v_{2}\,,\qquad
{\xi}_{0}\frak{Z}_{2}  =  -\frac{\Sigma}{\Delta}v_{3}\,,\qquad
{\xi}_{0}\frak{Q}_{10}  =  \sqrt{\frac{\Sigma}{\Delta}}v_{4}\,.
\end{eqnarray}
Similarly, for $T^1$, we have again 6 equations, namely
\begin{eqnarray} \label{Q-T-2}
  {\xi}_{0}(\frak{Q}_{3}+\frak{Q}_{4}) & = &
  -\sqrt{\frac{\Sigma}{\Delta}}v_{1} + {\xi}_{1}\frak{q}_{1}\,,\qquad
  {\xi}_{0}\frak{Y}_{3} = \frac{\Sigma}{\Delta}v_{2}\,,\qquad
  {\xi}_{0}\frak{Y}_{4} = \frac{\Sigma}{\Delta}v_{3}\,,\cr
{\xi}_{0}\frak{Z}_{3} & = & -\frac{\Sigma}{\Delta}v_{2}\,,\qquad
{\xi}_{0}\frak{Z}_{4}  =  -\frac{\Sigma}{\Delta}v_{3}\,,\qquad
{\xi}_{0}\frak{Q}_{11}  =  \sqrt{\frac{\Sigma}{\Delta}}v_{4}\,.
\end{eqnarray}
For $T^2$, one equation vanishes and three are redundant, since
contained in (\ref{Q-T-1}) or (\ref{Q-T-2}). Thus, we find the 2
equations
\begin{equation}\label{Q-T-3}
  {\xi}_{0}(\frak{Q}_{6}+\frak{Q}_{8}) =
  \sqrt{\frac{\Sigma}{\Delta}}v_{5} +
  {\xi}_{1}\frak{q}_{1}\,,\qquad{\xi}_{0}(\frak{Q}_{9}-\frak{Q}_{7}) =
  -\sqrt{\frac{\Sigma}{\Delta}}v_{5} + {\xi}_{1}\frak{q}_{0} \,.
\end{equation}
Eventually, for $T^3$, one equation vanishes and the rest is
redundant.

\subsection{Solving the master equation}

So far, we have 6+6+2=14 equations in (\ref{Q-T-1}), (\ref{Q-T-2}),
and (\ref{Q-T-3}) for the 12+4+4=20 functions $\frak{Q,Y,Z}$. In other
words, for making the system of equations well determined, we have to
pick 6 conditions. Two of them are obvious. We put $\frak{q}_0=
\frak{q}_1=0$, i.e., see (\ref{abbr}),
\begin{equation}\label{2cons}
  \frak{Q}_{1}-\frak{Q}_{5}- 2\frak{Q}_{9}=0\,,\qquad
  \frak{Q}_{0}-\frak{Q}_{4}-2\frak{Q}_{8}=0\,.
\end{equation}
The remaining 4 conditions can be selected from those equations in
(\ref{Q-T-1}), (\ref{Q-T-2}), and (\ref{Q-T-3}) in which sums of
$\frak{Q}$'s enter. We choose
\begin{equation}
\frak{Q}_{2} =  \frak{Q}_{3} = \frak{Q}_{8} = \frak{Q}_{9}  =  0\,.
\end{equation}
Then, according to (\ref{2cons}),
\begin{equation}
  \frak{Q}_{1}=\frak{Q}_{5}\qquad\hbox{and}\qquad
  \frak{Q}_{0}=\frak{Q}_{4}\,.
\end{equation}

Now, from (\ref{Q-T-1}), (\ref{Q-T-2}), and (\ref{Q-T-3}), it is
simple to read off the nonvanishing members of the $\frak{Q}$'s,
$\frak{Y}$'s, and $\frak{Z}$'s:
\begin{eqnarray}\label{haha}
 \frak{Q}_{0}& =& \frak{Q}_{1} = \frak{Q}_{4} = \frak{Q}_{5}  =
  -\frac{v_{1}}{{\xi}_{0}}\sqrt{\frac{\Sigma}{\Delta}}\,,\quad
  \frak{Q}_{6} = \frak{Q}_{7} =
  \frac{v_{5}}{{\xi}_{0}}\sqrt{\frac{\Sigma}{\Delta}}\,,\quad
  \frak{Q}_{10} = \frak{Q}_{11} =
  \frac{v_{4}}{{\xi}_{0}}\sqrt{\frac{\Sigma}{\Delta}}\,,
\cr
\frak{Y}_{1}& =& \frak{Y}_{3} = -\frak{Z}_{1} = -\frak{Z}_{3}  =
\frac{v_{2}}{{\xi}_{0}}{\frac{\Sigma}{\Delta}}\,,\qquad \frak{Y}_{2} =
\frak{Y}_{4} = -\frak{Z}_{2} = -\frak{Z}_{4}  =
  \frac{v_{3}}{{\xi}_{0}}{\frac{\Sigma}{\Delta}}\,.
\end{eqnarray}
If we substitute (\ref{haha}) into (\ref{Q_functions}) and
(\ref{1-forms}), the nonmetricity matrix turns out to be
\begin{eqnarray}
  Q^{\alpha\beta} & = &
  \frac{1}{{\xi}_{0}}\frac{\Sigma}{\Delta}\pmatrix{ 0 & 0 &
    v_{2}({\vartheta}^{0}-{\vartheta}^{1}) &
    v_{3}({\vartheta}^{0}-{\vartheta}^{1})\cr \bullet & 0 &
    v_{2}({\vartheta}^{0}-{\vartheta}^{1}) &
    v_{3}({\vartheta}^{0}-{\vartheta}^{1})\cr \bullet & \bullet & 0 &
    0 \cr \bullet & \bullet & \bullet & 0 }\cr & & \cr &
  &\hspace{-40pt} +
  \frac{1}{{\xi}_{0}}\sqrt{\frac{\Sigma}{\Delta}}\pmatrix{
    -v_{1}({\vartheta}^{0}+{\vartheta}^{1}) & 0 &
    v_{5}{\vartheta}^{2}+v_{4}{\vartheta}^{3} &
    -v_{4}{\vartheta}^{2}+v_{5}{\vartheta}^{3} \cr \bullet &
    -v_{1}({\vartheta}^{0}+{\vartheta}^{1}) &
    v_{5}{\vartheta}^{2}+v_{4}{\vartheta}^{3} &
    -v_{4}{\vartheta}^{2}+v_{5}{\vartheta}^{3} \cr \bullet &\bullet &
    0 & 0 \cr \bullet&\bullet&\bullet & 0 }\,.
\label{Q_matrix}
\end{eqnarray}

\subsection{Irreducible decomposition of the generated nonmetricity}

In order to recognize the structure of the nonmetricity, we decompose
it irreducibly in accordance with the scheme (\ref{deco4}) to
(\ref{deco1}):
\begin{eqnarray}
^{(1)}Q^{\alpha\beta} & = &
\frac{2v_{2}}{3{\xi_{0}}}\frac{\Sigma}{\Delta}\pmatrix{
-{\vartheta}^{2} & -{\vartheta}^{2} &
{\vartheta}^{0}-{\vartheta}^{1} & 0 \cr \bullet & -{\vartheta}^{2}
& {\vartheta}^{0}-{\vartheta}^{1} & 0 \cr \bullet & \bullet & 0 &
0 \cr \bullet & \bullet & \bullet & 0 } +
\frac{2v_{3}}{3{\xi_{0}}}\frac{\Sigma}{\Delta}\pmatrix{
-{\vartheta}^{3} & -{\vartheta}^{3} & 0 &
{\vartheta}^{0}-{\vartheta}^{1} \cr \bullet & -{\vartheta}^{3} & 0
& {\vartheta}^{0}-{\vartheta}^{1} \cr \bullet & \bullet & 0 & 0
\cr \bullet & \bullet & \bullet & 0 }\cr & & \cr
& &
+\frac{v_{1}}{9{\xi_{0}}}\sqrt{\frac{\Sigma}{\Delta}}\pmatrix{
-6{\vartheta}^{0}-4{\vartheta}^{1} &
4({\vartheta}^{0}+{\vartheta}^{1}) & {\vartheta}^{2} &
{\vartheta}^{3} \cr \bullet & -4{\vartheta}^{0}-6{\vartheta}^{1} &
{\vartheta}^{2}& {\vartheta}^{3} \cr \bullet & \bullet &
-({\vartheta}^{0}-{\vartheta}^{1}) & 0 \cr \bullet & \bullet &
\bullet & -({\vartheta}^{0}-{\vartheta}^{1}) }\cr & & \cr
&& + \frac{v_{5}}{9{\xi_{0}}}\sqrt{\frac{\Sigma}{\Delta}}\pmatrix{
-6{\vartheta}^{0}+2{\vartheta}^{1} &
-2({\vartheta}^{0}+{\vartheta}^{1}) & 4{\vartheta}^{2} &
4{\vartheta}^{3} \cr \bullet & 2{\vartheta}^{0}-6{\vartheta}^{1} &
4{\vartheta}^{2}& 4{\vartheta}^{3} \cr \bullet & \bullet &
-4({\vartheta}^{0}-{\vartheta}^{1}) & 0 \cr \bullet & \bullet &
\bullet & -4({\vartheta}^{0}-{\vartheta}^{1}) }\,,\cr & & \cr & & \cr
^{(2)}Q^{\alpha\beta} & = &
\frac{v_{2}}{3{\xi_{0}}}\frac{\Sigma}{\Delta}\pmatrix{
2{\vartheta}^{2} & 2{\vartheta}^{2} &
{\vartheta}^{0}-{\vartheta}^{1} & 0 \cr \bullet & 2{\vartheta}^{2}
& {\vartheta}^{0}-{\vartheta}^{1} & 0 \cr \bullet & \bullet & 0 &
0\cr \bullet & \bullet & \bullet & 0 }
+\frac{v_{3}}{3{\xi_{0}}}\frac{\Sigma}{\Delta}\pmatrix{
2{\vartheta}^{3} & 2{\vartheta}^{3} & 0 &
{\vartheta}^{0}-{\vartheta}^{1}\cr \bullet & 2{\vartheta}^{3} & 0
& {\vartheta}^{0}-{\vartheta}^{1}\cr \bullet & \bullet & 0 & 0\cr
\bullet & \bullet & \bullet & 0 }\,,\cr &&\cr &&\cr
^{(3)}Q^{\alpha\beta} & = &
-\frac{1}{9{\xi}_{0}}(v_{1}-2v_{5})\sqrt{\frac{\Sigma}{\Delta}}\pmatrix{
3{\vartheta}^{0}+{\vartheta}^{1} &
2({\vartheta}^{0}+{\vartheta}^{1}) & 2{\vartheta}^{2} &
2{\vartheta}^{3} \cr \bullet & {\vartheta}^{0}+3{\vartheta}^{1} &
2{\vartheta}^{2} & 2{\vartheta}^{3}\cr \bullet & \bullet &
{\vartheta}^{0}-{\vartheta}^{1} & 0 \cr \bullet & \bullet &
\bullet & {\vartheta}^{0}-{\vartheta}^{1} }\, , \cr
& & \cr
^{(4)}Q^{\alpha\beta} & = & 0\, .
\end{eqnarray}

For the 1-form triplet we find
\begin{eqnarray}\label{triplet}
  {\Lambda} & = &
  -\frac{v_{5}{\Sigma}}{{\xi}_{0}r^2}\sqrt{\frac{\Sigma}
    {\Delta}}({\vartheta}^{0}-{\vartheta}^{1})\,,\quad Q = 0\,,\quad
  T = -\frac{v_{5}{\Sigma}}{r^2}\sqrt{\frac{\Sigma}
    {\Delta}}({\vartheta}^{0}-{\vartheta}^{1})\,.
\end{eqnarray}
Remember that the function $v_5$ carries here an $L_0$ instead of the
original $M$, see (\ref{relax}). Because of (\ref{triplet}), we
have
\begin{equation}\label{prop}
 T= {\xi}_{0} {\Lambda}\,.
\end{equation}
This is a special case of (\ref{traceansatz}), compare also the
discussion in Heinicke et al.\cite{HBH_2005} where (\ref{traceansatz})
was used in the context of spherically symmetric exact solutions.

Besides $Q=0$, we also have $^{(3)}T^\alpha=0$, see
(\ref{axitor=0}). Moreover, from (\ref{prop}) we read off
$\xi_1=0$. Then a comparison with (\ref{eq19}) shows that
\begin{equation}\label{xi0=1/2}
\xi_0=1/2
\end{equation}
yields a particular simple connection. Thus, we adopt (\ref{xi0=1/2}).

Eventually we put the cosmological constant in the MAG Lagrangian
$\lambda_0$ equal to the corresponding Einsteinian cosmological
constant of our seed solution, i.e., $\lambda_0=\lambda$.

\section{Display of the solution}\label{displaysol}

Our new solution is given in terms of the coframe $\vta^\a$ in
(\ref{COFRAME}) [with the structure functions (\ref{DSF}) to
(\ref{DSF'})], of the metric $g$ in (\ref{anhmetric}), of the torsion
[see (\ref{TORSION-U4})]
\begin{eqnarray}
  T^{0} & = &
  T^{1}=-\frac{{L_0}(r^2-a^2{\cos^2\theta})}{{\Sigma}\sqrt{{\Delta}
      {\Sigma}}} {\vartheta}^{01} -\frac{{L_0}a^{2}
    r{\sin\theta}{\cos\theta}} {\Sigma^2}\sqrt{\frac{F}
    {\Delta}}({\vartheta}^{02} -{\vartheta}^{12})\, , \cr & & +
  \frac{{L_0}ar^2{\sin\theta}}{{\Delta}
    {\Sigma}}\sqrt{\frac{F}{\Sigma}}({\vartheta}^{03}
  -{\vartheta}^{13}) - \frac{2{L_0}ar\,{\cos\theta}}{{\Sigma}
    \sqrt{{\Delta}{\Sigma}}}{\vartheta}^{23}\, , \cr T^{2} & = &
  \frac{{L_0}r^2}{{\Sigma}\sqrt{{\Delta}{\Sigma}}}
  ({\vartheta}^{02}-{\vartheta}^{12})+ \frac{{L_0}ar\,{\cos\theta}}{{\Sigma}
    \sqrt{{\Delta}{\Sigma}}}({\vartheta}^{03}-{\vartheta}^{13})\, ,
  \cr T^{3} & = & - \frac{{L_0}ar\,{\cos\theta}}{{\Sigma}
    \sqrt{{\Delta}{\Sigma}}}({\vartheta}^{02}-{\vartheta}^{12}) +
  \frac{{L_0}r^2}{{\Sigma}\sqrt{{\Delta}
      {\Sigma}}}({\vartheta}^{03}-{\vartheta}^{13}) \, ,
\end{eqnarray}
and of the nonmetricity [see (\ref{Q_matrix})]
\begin{eqnarray}\label{nonmetr}
  Q^{\alpha\beta} & = &
  \frac{2L_{0}ar\,{\sin\theta}}{{\Delta}{\Sigma}}\sqrt{\frac{F}{\Sigma}}
  \pmatrix{ 0 & 0 & -a{\cos\theta}({\vartheta}^{0}-{\vartheta}^{1}) &
    r({\vartheta}^{0}-{\vartheta}^{1})\cr \bullet & 0 &
    -a{\cos\theta}({\vartheta}^{0}-{\vartheta}^{1}) &
    r({\vartheta}^{0}-{\vartheta}^{1})\cr \bullet & \bullet & 0 & 0\cr
    \bullet & \bullet & \bullet & 0 }\cr & &\cr & &\cr & & -
  \frac{2L_{0}}{{\Sigma}\sqrt{{\Delta}{\Sigma}}}
  \pmatrix{\O & 0 &
    -(r^2{\vartheta}^{2}+ar{\cos\theta}{\vartheta}^{3}) &
    ar{\cos\theta}{\vartheta}^{2}-r^2{\vartheta}^{3}\cr \bullet &
    \O & -(r^2{\vartheta}^{2}+ar{\cos\theta}{\vartheta}^{3}) &
    ar{\cos\theta}{\vartheta}^{2}-r^2{\vartheta}^{3}\cr \bullet &
    \bullet & 0 & 0\cr \bullet & \bullet & \bullet & 0 },
\end{eqnarray}
with $\O:= (r^2-a^2{\cos^2\theta})({\vartheta}^{0}+ {\vartheta}^{1})$.
This completes our solution belonging to the Lagrangians (\ref{VI}).

The explicit verification that the field equations are fulfilled,
indeed, is still a delicate task. We did it by means of Hearn's
computer algebra system REDUCE together with Schr\"ufer's EXCALC
package, see also Socorro et al.\cite{Socorro}, Heinicke et
al.\cite{Grab}, and Ref.\cite{Birkbook}.

In studying the properties of our solution, certainly the computation
of the curvature will provide some insight. In \ref{Solcurv} we
collected all the corresponding formulas. It becomes immediately clear
that our solution is far from being trivial. It rather displays a
fairly complicated structure. In order to get some insight, we will
display first a typical component of the rotational curvature, namely a
component of the Weyl curvature,
\begin{eqnarray}\label{W03}
  ^{(1)}W^{03}= \frac{L_{0}^2ar\,{\sin\theta}(r^2-a^2{\cos^2\theta})}
  {2{\Delta}{\Sigma}^3}\sqrt{\frac{F}{\Delta}}
  \left(r\vta^{01}-a\,{\cos\theta}\vta^{23} \right)\,,
\end{eqnarray}
and such a component of the strain curvature that is not too
complicated,
\begin{eqnarray}\label{Z}
  ^{(2)}Z^{00} =
  \frac{ML_{0}ar\,{\cos\theta}}{{\Delta}{\Sigma}^3}
  (3r^2-a^2{\cos^2\theta}) \,\vta^{23}\,.
\end{eqnarray}
The other pieces are listed in \ref{SolcurvW} and \ref{SolcurvZ},
respectively.

Alternatively, instead of torsion and nonmetricity, we could display
the connection of our solution. It can be read off from
(\ref{eq19}). Since $Q=0$, we have 
\begin{equation}\label{eq19'}
  {\Gamma}_{\alpha\beta} = \widetilde{\Gamma}_{\alpha\beta} + {1\over
    2}Q_{\alpha\beta}=\widetilde{\Gamma}_{\alpha\beta} + {1\over
    2}{\not\! Q}_{\alpha\beta}\,.
\end{equation}
To the general relativistic Levi-Civita connection
$\widetilde{\Gamma}_{\alpha\beta}$, we have to add half of the
nonmetricity (\ref{nonmetr}).


\section*{Acknowledgments} 
One of us (P.B.) would like to thank Metin G\"urses (Ankara) for his
advice and his encouragement to apply prolongation methods in MAG. We
are grateful to Christian Heinicke (Cologne) for discussions on
metric-affine geometry.

\appendix

\section{Irreducible decompositions of the geometrical field
strengths}\label{appdec}

At each point of spacetime, we have invariance under the linear group
$GL(4,R)$. Therefore we can decompose nonmetricity, torsion, and
curvature irreducibly under this group. Moreover, since a metric is
defined locally, we can decompose these quantities even finer, namely
with respect to the Lorentz group $SO(1,3)$. This is what we
will list here, for more details and references to the original
literature, see Ref.\cite{PRs}.

\subsection{Decomposition of the nonmetricity}

The nonmetricity $Q_{\alpha\beta}$ can be decomposed into four pieces.
We have to recapitulate some of these features. In four dimensions, as
a symmetric tensor-valued $1$-form, the nonmetricity has 40
independent components. Two vector-like pieces can be easily
identified. Firstly, the Weyl covector $Q:=Q_\a{}^\a/4$ can be
extracted by tracing $Q_{\alpha\beta}$. The remaining tracefree part
of the nonmetricity $\not\!\!Q_{\alpha\beta}$ contains a second
vector-like piece represented by the $1$-form ${\Lambda}$:
\begin{equation}\label{Lambda}
\Lambda := \left(e^\beta \rfloor \qslash_{\alpha\beta}\right) \,
\wedge \vartheta^\alpha \,.
\end{equation}
The 2-form\cite{HBH_2005}
\begin{equation}
 P _{\alpha} := \qslash_{\alpha\beta} \wedge
  \vartheta^\beta - \frac{1}{3} \, \vartheta_\alpha \wedge \Lambda\,,
\label{P_alpha}
\end{equation}
obeys the $4+4$ constraints
\begin{equation}
 P^{\alpha}\wedge {\vartheta}_{\alpha}  =  0\,,\qquad
e_{\alpha}\rfloor P^{\alpha}  = 0\,.
\end{equation}
Accordingly, $P^\alpha$ carries $24-4-4=16$ independent components,
and it is related to a further irreducible piece of $Q_{\alpha\beta}$.
Explicitly, we have
\begin{eqnarray}\label{deco4}
{}^{(4)}Q_{\alpha\beta} & := & Q \, g_{\alpha\beta} \,,\quad\\
  {}^{(3)}Q_{\alpha\beta} & := & \frac{4}{9} \left(
    \vartheta_{(\alpha} \,e_{\beta )}\rfloor - \frac{1}{4} \,
    g_{\alpha\beta} \right)\Lambda \,,\label{3Q} \\
  {}^{(2)}Q_{\alpha\beta} & := & -\frac{2}{3} \, e_{(\alpha} \rfloor
   P _{\beta)}\label{Q2} \,, \\
{}^{(1)}Q_{\alpha\beta} &:=& Q_{\alpha\beta} -
  {}^{(2)}Q_{\alpha\beta} - {}^{(3)}Q_{\alpha\beta} -
  {}^{(4)}Q_{\alpha\beta} \,.\label{deco1}
\end{eqnarray}

\subsection{Decomposition of the torsion}

The torsion $T^\a$ can be decomposed irreducibly into three
independent pieces, the totally antisymmetric axial part
$^{(3)}T^{\alpha}$, its trace $^{(2)}T^{\alpha}$, and the tracefree
symmetric tensor part $^{(1)}T^{\alpha}$. They read, respectively,
\begin{eqnarray}\label{tor3}
  ^{(3)}T^{\alpha} & := & \frac{1}{3}e^{\alpha}\rfloor
  ({\vartheta}^{\mu}\wedge T_{\mu})\, , \\ & & \cr\label{tor2}
  ^{(2)}T^{\alpha} & := & \frac{1}{3}{\vartheta}^{\alpha}\wedge T\quad
  {\rm with}\quad T:=e_{\mu}\rfloor T^{\mu}\, , \\ & & \cr
  ^{(1)}T^{\alpha} & := & T^{\alpha} - {}^{(2)}T^{\alpha} -
  {}^{(3)}T^{\alpha}\, .
\label{decomp_torsion}
\end{eqnarray}

\subsection{Decomposition of the strain = shear $\oplus$ dilation
curvature $Z_{\alpha\beta}$}\label{straincurv}

From the strain curvature $Z_{\a\b}:=R_{(\a\b)}$ we can split off the
dilation curvature $Z:=R_\a{}^\a$, see (\ref{curvdef}). The
(tracefree) shear curvature $\zslash_{\alpha\beta}$ can be cut into
different pieces by contraction with $e_\alpha$, transvecting with
$\vartheta^{\alpha}$, and by ``hodge''-ing the corresponding
expressions:
\begin{equation}
\label{ZDY}\zslash_\alpha:=e^\beta\rfloor \zslash_{\alpha\beta},
\qquad \hat\Delta:={1\over 2}\,\vartheta^\alpha
\wedge\zslash_\alpha,\qquad Y_\alpha:=\,^\star
(\zslash_{\alpha\beta}\wedge \vartheta^\beta)\,.
\end{equation}
Subsequently we can subtract out traces:
\begin{equation}
  \Xi_\alpha:= \zslash_\alpha-{1\over2}e_\alpha\rfloor
  (\vartheta^\gamma \wedge\zslash_\gamma), \qquad \qquad
  \Upsilon_\alpha:= Y_\alpha- {1\over
    2}\,e_\alpha\rfloor(\vartheta^\gamma\wedge Y_\gamma)\,.
\end{equation}
The irreducible pieces may then be written as
\begin{eqnarray}
^{(2)}\!Z_{\alpha\beta}&:=& - {1\over 2}\,^\star(\vartheta_{(\alpha}
\wedge\Upsilon_{\beta)})\,,\\
\label{Z3de}^{(3)}\!Z_{\alpha\beta}&:=&\hspace{8pt}{1\over
  3}\;\left( 2\,\vartheta_{(\alpha}\wedge e_{\beta)}\rfloor-\,
  g_{\alpha\beta}\right)\hat\Delta\,,\\ ^{(4)}\!
Z_{\alpha\beta}&:=&\hspace{8pt} {1\over 4}\;g_{\alpha\beta}\,Z\,,\\
^{(5)}\! Z_{\alpha\beta}&:=& \hspace{8pt}{1\over 2}\;
\vartheta_{(\alpha}\wedge \Xi_{\beta)}\,,\\
^{(1)}\!Z_{\alpha\beta}&:=& \hspace{8pt} Z_{\alpha\beta}-
\,^{(2)}\!Z_{\alpha\beta}- \,^{(3)}\!Z_{\alpha\beta}
-\,^{(4)}\!Z_{\alpha\beta}- \,^{(5)}\!Z_{\alpha\beta}\,.
\end{eqnarray}

\subsection{Decomposition of the rotational curvature}\label{rotcurv}

The rotational curvature $W^{\alpha\beta}:=R^{[\alpha\beta]}$ is a sum
of six irreducible pieces,
\begin{equation}
W^{\alpha\beta} = \sum\limits_{k=1}^{6}\, {^{(k)}W^{\alpha\beta}}
\end{equation}
that can be parameterized by using the following four vector-valued
1-forms $W^{\alpha}$, $X^{\alpha}$, ${\Phi}^{\alpha}$, and
${\Psi}^{\alpha}$:
\begin{eqnarray}
 W^{\alpha} & := & e_{\beta}\rfloor W^{\alpha\beta}\,,\\
 & & \cr
 X^{\alpha} & := & ^{\star}(W^{\beta\alpha}\wedge
 {\vartheta}_{\beta})\,,\\
 & & \cr
 {\Phi}_{\alpha} & := &
 W_{\alpha}-\frac{1}{4}W{\vartheta}_{\alpha}-\frac{1}{2}e_{\alpha}\rfloor
 ({\vartheta}^{\mu}\wedge W_{\mu})\,,\\
 & & \cr
 {\Psi}_{\alpha} & := & X_{\alpha}-\frac{1}{4}X{\vartheta}_{\alpha}
 -\frac{1}{2}e_{\alpha}\rfloor ({\vartheta}^{\mu}\wedge X_{\mu})\,.
\end{eqnarray}
The 0-forms $W$ and $X$ are the contractions of the corresponding
1-forms, i.e.
\begin{equation}
  W = e_{\mu}\rfloor W^{\mu}\,,\qquad X = e_{\mu}\rfloor
  X^{\mu}\,,
\end{equation}
whereas the contractions of ${\Phi}_{\alpha}$ and
${\Psi}_{\alpha}$ vanish identically, i.e.,
\begin{equation}
  e_{\mu}\rfloor {\Phi}^{\mu} = 0\,, \qquad e_{\mu}\rfloor
  {\Psi}^{\mu}  =  0\,.
\end{equation}
Furthermore, we find
\begin{equation}
 {\Phi}_{\alpha}\wedge {\vartheta}^{\alpha}  =  0\,.
\end{equation}

For the irreducible pieces of the rotational curvature in terms of
these auxiliary 1-forms there results
\begin{eqnarray}\label{wewe}
  ^{(2)}W^{\alpha\beta} & = & - {}^{\star}({\vartheta}^{[
    \alpha}\wedge {\Psi}^{\beta ]} )\, , \\ & & \cr
  ^{(3)}W^{\alpha\beta} & = & -\frac{1}{12} {^{\star}(X\wedge
    {\vartheta}^{\alpha}\wedge {\vartheta}^{\beta} )}\, ,\\ & & \cr
  ^{(4)}W^{\alpha\beta} & = & -\frac{1}{2}{\vartheta}^{[ \alpha}\wedge
  {\Phi}^{\beta ]}\, , \\ & & \cr ^{(5)}W^{\alpha\beta} & = &
  -\frac{1}{2}{\vartheta}^{[ \alpha}\wedge e^{\beta ]}\rfloor
  ({\vartheta}^{\mu}\wedge W_{\mu})\, , \\ & & \cr
  ^{(6)}W^{\alpha\beta} & = & -\frac{W}{12}{\vartheta}^{\alpha}\wedge
  {\vartheta}^{\beta}\, ,\label{6W} \\ & & \cr ^{(1)}W^{\alpha\beta} & = &
  W^{\alpha\beta}-\sum\limits_{n=2}^{6} {^{(n)}W^{\alpha\beta}}\,
  .\label{w1w1}
\end{eqnarray}

\section{Decomposing the curvature of our solution}\label{Solcurv}

\subsection{Irreducible rotational curvature $R^{[\alpha\beta ]}$}
\label{SolcurvW}

To characterize the irreducible pieces of the rotational curvature it
is of advantage to introduce the following four curvature structure
functions ${\Phi}_{1},\cdots {\Phi}_{4}$, all depending on the
non-ignorable coordinates $(r,\theta )$ as follows:
\begin{eqnarray}\label{rotcurv'}
  {\Phi}_{1} & := &
  \frac{L_{0}^2a^2r{\sin\theta}{\cos\theta}(r^2-a^2{\cos^2\theta})}
   {2{\Delta}{\Sigma}^3}\sqrt{\frac{F}{\Delta}}\, ,\cr
& & \cr {\Phi}_{2} & := &
  \frac{L_{0}^2ar^2{\sin\theta}(r^2-a^2{\cos^2\theta})}
   {2{\Delta}{\Sigma}^3}\sqrt{\frac{F}{\Delta}}\, ,\cr
& & \cr {\Phi}_{3} & := &
  \frac{L_{0}^2ar{\cos\theta}(r^2-a^2{\cos^2\theta})}
   {{\Delta}{\Sigma}^3}\, , \cr
& & \cr {\Phi}_{4} & := &
  \frac{L_{0}^2r^2(r^2-a^2{\cos^2\theta})}
   {{\Delta}{\Sigma}^3}\, .
\end{eqnarray}
In this way, the curvature $R^{[\alpha\beta ]}$ has a relative simple
appearance. Observe that these functions are related algebraically
according to
\begin{eqnarray}
 {\Phi}_{1}r-{\Phi}_{2}a{\cos\theta} & = & 0\, ,\cr
 {\Phi}_{3}r-{\Phi}_{4}a{\cos\theta} & = & 0\, .
\end{eqnarray}
The corresponding relations for the torsion functions
(\ref{TORSION-U4}) read
\begin{eqnarray}
v_{2}r+v_{3}a{\cos\theta} & = & 0\, ,\cr
(v_{1}-v_{5})r+v_{4}a{\cos\theta} & = & 0\, .
\end{eqnarray}
Because of
\begin{equation}
\frac{{\Phi}_{1}}{{\Phi}_{2}} = \frac{{\Phi}_{3}}{{\Phi}_{4}}\quad
{\rm and}\quad \frac{v_{2}}{v_{1}-v_{5}}=\frac{v_{3}}{v_{4}}\,,
\end{equation}
these functions are not functionally independent. It is also
remarkable that the $\Phi$'s, and thus also the rotational curvature
$W^{\a\b}$, depend on the mass only via the function $\Delta$.
However, the constant $L_0^2$ appears in all $\Phi$'s as
proportionality constant.

For the (generalized) Weyl curvature $^{(1)}W^{\alpha\beta}$ ({\bf
WEYL}) we find
\begin{eqnarray}
^{(1)}W^{\alpha\beta} & = & ^{(1)}{\widetilde W}^{\alpha\beta}\cr
& & \cr & & + {\Phi}_{1}\pmatrix{0 &
-({\vartheta}^{02}+{\vartheta}^{12}) &
    -{\vartheta}^{01} & -{\vartheta}^{23} \cr
\diamond & 0 & -{\vartheta}^{01} & {\vartheta}^{23} \cr \diamond &
\diamond & 0 & {\vartheta}^{03}+{\vartheta}^{13} \cr \diamond &
\diamond & \diamond & 0} \cr & & \cr & & + {\Phi}_{2}\pmatrix{0 &
{\vartheta}^{03}+{\vartheta}^{13} &
    -{\vartheta}^{23} & {\vartheta}^{01} \cr
\diamond & 0 & -{\vartheta}^{23} & -{\vartheta}^{01} \cr \diamond
& \diamond & 0 & {\vartheta}^{02}+{\vartheta}^{12} \cr \diamond &
\diamond & \diamond & 0}\,,
\end{eqnarray}
where $^{(1)}{\widetilde W}^{\alpha\beta}$ denotes the Riemannian
part, cf.\ (\ref{WEYL-V4}).

The pair-commutator $^{(2)}W^{\alpha\beta}$ ({\bf PAIRCOM}) turns
out to be
\begin{eqnarray}
  ^{(2)}W^{\alpha\beta} & = & {\Phi}_{1} \pmatrix{0 &
    {\vartheta}^{02}+{\vartheta}^{12} &
    -{\vartheta}^{01} & {\vartheta}^{23} \cr
    \diamond & 0 & {\vartheta}^{01} & -{\vartheta}^{23} \cr \diamond &
    \diamond & 0 & {\vartheta}^{03}+{\vartheta}^{13} \cr \diamond &
    \diamond & \diamond & 0 }\cr & & \cr & + & {\Phi}_{2} \pmatrix{ 0
    & -({\vartheta}^{03}+{\vartheta}^{13}) &
    {\vartheta}^{23} & {\vartheta}^{01} \cr
    \diamond & 0 & -{\vartheta}^{23} & -{\vartheta}^{01} \cr \diamond
    & \diamond & 0 & {\vartheta}^{02} + {\vartheta}^{12} \cr \diamond
    & \diamond & \diamond & 0 } \cr & & \cr
  & + & {\Phi}_{3}\pmatrix{ 0 & 0 &
    -({\vartheta}^{03}+{\vartheta}^{13}) &
    {\vartheta}^{02}+{\vartheta}^{12} \cr \diamond & 0 &
    {\vartheta}^{03}+{\vartheta}^{13} &
    -({\vartheta}^{02}+{\vartheta}^{12}) \cr
    \diamond & \diamond & 0 & 0 \cr \diamond & \diamond & \diamond & 0
  }.
\end{eqnarray}
The pseudoscalar part of the curvature ({\bf PSCALAR}) vanishes
identically, i.e.,
\begin{equation}
^{(3)}W^{\alpha\beta} = 0\,.
\end{equation}
The tracefree symmetric Ricci $^{(4)}W^{\alpha\beta}$ ({\bf
RICSYMF}) turns out to be
\begin{eqnarray}
 ^{(4)}W^{\alpha\beta} & = & {\Phi}_{2}\pmatrix{
    0 & {\vartheta}^{03}+{\vartheta}^{13} & {\vartheta}^{23} &
  {\vartheta}^{01} \cr
\diamond & 0 & -{\vartheta}^{23} & -{\vartheta}^{01} \cr \diamond
& \diamond & 0 & -({\vartheta}^{02}+{\vartheta}^{12}) \cr \diamond
& \diamond & \diamond & 0 } \cr & & \cr & + & {\Phi}_{1}\pmatrix{
    0 & -({\vartheta}^{02}+{\vartheta}^{12}) & -{\vartheta}^{01} &
  {\vartheta}^{23} \cr
\diamond & 0 & {\vartheta}^{01} & -{\vartheta}^{23} \cr \diamond &
\diamond & 0 & -({\vartheta}^{03}+{\vartheta}^{13}) \cr \diamond &
\diamond & \diamond & 0 } \cr & & \cr & + & {\Phi}_{4} \pmatrix{
    0 & 0 & -({\vartheta}^{02}+{\vartheta}^{12}) &
    -({\vartheta}^{03}+{\vartheta}^{13}) \cr
\diamond & 0 & {\vartheta}^{02}+{\vartheta}^{12} &
   {\vartheta}^{03}+{\vartheta}^{13} \cr
\diamond & \diamond & 0 & 0 \cr \diamond & \diamond & \diamond & 0
}\,,
\end{eqnarray}
and the antisymmetric Ricci $^{(5)}W^{\alpha\beta}$ ({\bf
RICANTI}) and the curvature scalar part $^{(6)}W^{\alpha\beta}$
({\bf SCALAR}) read, respectively,
\begin{eqnarray}
^{(5)}W^{\alpha\beta} & = & {\Phi}_{1} \pmatrix{ 0 &
{\vartheta}^{02}+{\vartheta}^{12} & -{\vartheta}^{01} &
-{\vartheta}^{23} \cr \diamond & 0 & {\vartheta}^{01} &
{\vartheta}^{23} \cr \diamond & \diamond & 0 & -({\vartheta}^{03}+
{\vartheta}^{13})\cr \diamond & \diamond & \diamond & 0 }\cr & &
\cr & & \cr & + & {\Phi}_{2} \pmatrix{ 0 &
-({\vartheta}^{03}+{\vartheta}^{13}) & -{\vartheta}^{23} &
{\vartheta}^{01} \cr \diamond & 0 & {\vartheta}^{23} &
-{\vartheta}^{01} \cr \diamond & \diamond & 0 &
-({\vartheta}^{02}+ {\vartheta}^{12})\cr \diamond & \diamond &
\diamond & 0 }\; {\rm and} \\ & & \cr ^{(6)}W^{\alpha\beta} & = &
-\frac{\lambda}{3}{\vartheta}^{\alpha\beta}\,.\label{wscalar}
\end{eqnarray}
Note that (\ref{wscalar}), because of (\ref{6W}), is consistent with
(\ref{Wconst}).

\subsection{Decomposition of the strain curvature $Z^{\alpha\beta}$}
\label{SolcurvZ}

In view of the complexity of the irreducible piece
$^{(1)}Z^{\alpha\beta}$ we will not display it in terms of matrices
but give the result just in terms of components,
\begin{footnotesize}
\begin{eqnarray}
^{(1)}Z^{00} & = & \left[
\frac{L_{0}r(r^2-3a^2{\cos^2\theta})}{{\Delta}{\Sigma}^3}(Mr-{\Delta})-
\frac{L_{0}r(r^2-a^2{\cos^2\theta})}{{\Delta}{\Sigma}^2}\left( 1 -
\frac{\lambda}{3}(a^2 + 2r^2)\right) \right]{\vartheta}^{01}\cr &
& \cr & & +
\frac{6L_{0}a^2r^2{\sin\theta}{\cos\theta}}{{\Sigma}^3}\sqrt{\frac{F}{\Delta}}{\vartheta}^{02}
+\frac{2L_{0}a^3r{\sin\theta}{\cos^2\theta}}{{\Sigma}^3}\sqrt{\frac{F}{\Delta}}
({\vartheta}^{03}-{\vartheta}^{12})\cr & &\cr & & +
\frac{2L_{0}ar^3{\sin\theta}}{\Sigma^3}\sqrt{\frac{F}{\Delta}}{\vartheta}^{13}
-\frac{L_{0}a{\cos\theta}(3r^2-a^2{\cos^2\theta})}{{\Delta}\Sigma^3}(Mr+2{\Delta}){\vartheta}^{23}\,,\cr
& & \cr ^{(1)}Z^{01} & = & -\frac{L_{0}}{\Delta\Sigma^{3}}\left[
M(2r^4-a^2r^2{\cos^2\theta}+a^4{\cos^4\theta})-2r(r^2-a^2{\cos^2\theta})\left(
\frac{\lambda}{3}\Sigma^2 +
a^2{\sin^2\theta}F\right)\right]{\vartheta}^{01}\cr & & \cr & &
-\frac{L_{0}a^2r^2{\sin\theta}{\cos\theta}}{{\Sigma}^3}\sqrt{\frac{F}{\Delta}}
({\vartheta}^{02}+{\vartheta}^{12}) +
\frac{2L_{0}ar^3{\sin\theta}}{{\Sigma}^3}\sqrt{\frac{F}{\Delta}}
(2{\vartheta}^{03}+{\vartheta}^{13})\cr & & \cr & &
-\frac{L_{0}a^3r{\sin\theta}{\cos^2{\theta}}}{{\Sigma}^3}\sqrt{\frac{F}{\Delta}}
({\vartheta}^{03}+3{\vartheta}^{13})
-\frac{L_{0}ar{\cos\theta}}{\Delta\Sigma^3}\left[
M(3r^2-a^2{\cos^2\theta})+4{\Delta}r\right]{\vartheta}^{23}\,,\cr & &
\cr
 ^{(1)}Z^{02} & = &
-\frac{L_{0}a^2r^2{\sin\theta}{\cos\theta}}{{\Sigma}^3}\sqrt{\frac{F}{\Delta}}
{\vartheta}^{01} \cr & & + \frac{L_{0}}{2\Delta\Sigma^3}\left[
M(2r^4-a^2r^2{\cos^2\theta}+a^4{\cos^4\theta})
-2r(r^2-a^2{\cos^2\theta})\left( \frac{\lambda}{3}{\Sigma}^2 +
a^2{\sin^2\theta}F)\right)\right]{\vartheta}^{02}\cr & & \cr & & +
\frac{L_{0}a{\cos\theta}(3r^2-a^2{\cos^2\theta})}{2{\Delta}{\Sigma}^3}(Mr+2{\Delta})
{\vartheta}^{03}\cr &  & \cr & &
\left[\frac{L_{0}r(r^2-3a^2{\cos^2\theta})}{2\Delta\Sigma^3}(Mr-{\Delta})
-\frac{L_{0}r(r^2-a^2{\cos^2\theta})}{2\Delta\Sigma^2}
\left(1-\frac{\lambda}{3}(a^2+2r^2)\right)\right]{\vartheta}^{12}\cr
& &  \cr & & - \frac{L_{0}a{\cos{\theta}}}{2{\Delta}{\Sigma}^3}
\left[4r^2{\Delta} + Mr(3r^2-a^2{\cos^2 {\theta}})
\right]{\vartheta}^{13} +
\frac{L_{0}ar{\sin\theta}}{{\Sigma}^3}\sqrt{\frac{F}{\Delta}}({\Sigma}-3r^2)
{\vartheta}^{23}\,,\cr & &\cr
^{(1)}Z^{03} & = &
\frac{3L_{0}ar{\sin\theta}(3r^2-{\Sigma})}{{\Sigma}^3}\sqrt{\frac{F}{\Delta}}
{\vartheta}^{01} -
\frac{L_{0}a{\cos\theta}(3r^2-a^2{\cos^2\theta})}{2\Delta\Sigma^3}(Mr+2{\Delta})
{\vartheta}^{02}\cr & & \cr & & +
\frac{L_{0}}{2\Delta\Sigma^3}\left[
M(2r^4-a^2r^2{\cos^2\theta}+a^4{\cos^4\theta})
-2r(r^2-a^2{\cos^2\theta})\left( \frac{\lambda}{3}{\Sigma}^2 +
a^2{\sin^2\theta}F\right)\right]{\vartheta}^{03}\cr & & \cr & & +
\frac{L_{0}a{\cos\theta}}{2{\Delta}{\Sigma}^3} \left[
4r^2{\Delta}+Mr(3r^2-a^2{\cos^2\theta})\right]
     {\vartheta}^{12}\cr & & \cr
& & +
\frac{L_{0}r}{2\Delta\Sigma^3}\left[(r^2-3a^2{\cos^2\theta})(Mr-{\Delta})
-(r^2-a^2{\cos^2\theta})
\left(1-\frac{\lambda}{3}(a^2+2r^2)\right)\right]{\vartheta}^{13}\cr
& & \cr & & +
\frac{5L_{0}a^2r^2{\sin\theta}{\cos\theta}}{{\Sigma}^3}\sqrt{\frac{F}{\Delta}}
{\vartheta}^{23}\,, \cr & & \cr ^{(1)}Z^{11} & = &
\frac{L_{0}r}{\Delta\Sigma^3}\left[(r^2-3a^2{\cos^2\theta})(-Mr+{\Delta})
- {\Sigma}(r^2-a^2{\cos^2\theta})\left( 1 - \frac{\lambda}{3}(a^2
+ 2r^2)\right)\right]{\vartheta}^{01}\cr & & \cr & & +
\frac{2L_{0}ar^3{\sin\theta}}{{\Sigma}^3}\sqrt{\frac{F}{\Delta}}
(2{\vartheta}^{03}-{\vartheta}^{13})
+\frac{6L_{0}a^2r^2{\sin\theta}{\cos\theta}}{{\Sigma}^3}\sqrt{\frac{F}{\Delta}}
{\vartheta}^{12}\cr & &\cr & & -
\frac{L_{0}a{\cos\theta}(3r^2-a^2{\cos^2\theta})}{{\Delta}{\Sigma}^3}
(Mr+2{\Delta})\,, \cr & & \cr ^{(1)}Z^{12} & = &
-\frac{L_{0}a^2r^2{\sin\theta}{\cos\theta}}{{\Sigma}^3}\sqrt{\frac{F}{\Delta}}
{\vartheta}^{01} \cr & & \cr & & +
\frac{L_{0}r}{2\Delta\Sigma^3}\left[
(r^2-3a^2{\cos^2\theta})({\Delta}-Mr)
+(r^2-a^2{\cos^2\theta})\left(1- \frac{\lambda}{3}(a^2+2r^2)
\right)\right]{\vartheta}^{02}\cr & & \cr & & -
\frac{L_{0}}{2\Delta\Sigma^3}\left[M(2r^4-a^2r^2{\cos^2\theta}+a^4{\cos^4\theta})
-2r(r^2-a^2{\cos^2\theta}) \left(\frac{\lambda}{3}{\Sigma}^2 +
a^2{\sin^2\theta}F\right)\right]{\vartheta}^{12}\cr & & \cr & & -
\frac{L_{0}a{\cos{\theta}}}{2{\Delta}{\Sigma}^3}(3r^2-a^2{\cos^2
{\theta}})(Mr+{\Delta}){\vartheta}^{13} +
\frac{L_{0}ar{\sin\theta}}{{\Sigma}^3}\sqrt{\frac{F}{\Delta}}({\Sigma}-3r^2)
{\vartheta}^{23}\,,\cr & &\cr ^{(1)}Z^{13} & = &
\frac{3L_{0}ar{\sin\theta}(3r^2-{\Sigma})}{{\Sigma}^3}\sqrt{\frac{F}{\Delta}}
{\vartheta}^{01}
-\frac{L_{0}ar{\cos{\theta}}}{2{\Delta}{\Sigma}^3}\left[
4{\Delta}r + M(3r^2-a^2{\cos^2
{\theta}})\right]{\vartheta}^{02}\cr & & \cr & & -
\frac{L_{0}r}{2\Delta\Sigma^3}\left[
(r^2-3a^2{\cos^2\theta})(Mr-{\Delta}) - (r^2-a^2{\cos^2\theta})
\left(1-\frac{\lambda}{3}(a^2+2r^2)\right)
\right]{\vartheta}^{03}\cr & & \cr & & -
\frac{L_{0}}{2{\Delta}{\Sigma}^3}\left[M(2r^4-a^2r^2{\cos^2\theta}+a^4{\cos^4\theta})
-2r(r^2-a^2{\cos^2\theta}) \left(\frac{\lambda}{3}{\Sigma}^2 +
a^2{\sin^2\theta}F\right)\right]{\vartheta}^{13}\cr & & \cr & & +
\frac{L_{0}a{\cos{\theta}}}{2{\Delta}{\Sigma}^3}(3r^2-a^2{\cos^2\theta})(Mr+2{\Delta})
{\vartheta}^{12}+
\frac{5L_{0}a^2r^2{\sin\theta}{\cos\theta}}{{\Sigma}^3}\sqrt{\frac{F}{\Delta}}
{\vartheta}^{23}\,,
\cr & & \cr ^{(1)}Z^{22} & = &
\frac{2L_{0}a^2r^2{\sin\theta}{\cos\theta}}{{\Sigma}^3}\sqrt{\frac{F}{\Delta}}
({\vartheta}^{02}-{\vartheta}^{12})\,,\cr & & \cr ^{(1)}Z^{23} & = &
-\frac{L_{0}ar{\sin\theta}(3r^2-{\Sigma})}{{\Sigma}^3}\sqrt{\frac{F}{\Delta}}
({\vartheta}^{02}-{\vartheta}^{12})
-\frac{L_{0}a^2r^2{\sin\theta}{\cos\theta}}{{\Sigma}^3}\sqrt{\frac{F}{\Delta}}
({\vartheta}^{03}-{\vartheta}^{13})\,,\cr & & \cr ^{(1)}Z^{33} & = &
-\frac{2L_{0}ar{\sin\theta}(3r^2-{\Sigma})}{{\Sigma}^3}\sqrt{\frac{F}{\Delta}}
({\vartheta}^{03}-{\vartheta}^{13}) +
\frac{4L_{0}a^2r^2{\sin\theta}{\cos\theta}}{{\Sigma}^3}\sqrt{\frac{F}{\Delta}}
({\vartheta}^{02}-{\vartheta}^{12})\,,
\end{eqnarray}


\begin{equation}
  ^{(2)}Z^{\alpha\beta} =
  \frac{ML_{0}ar{\cos\theta}}{2{\Delta}{\Sigma}^3}(3r^2-a^2{\cos^2\theta})
  \pmatrix{
    2{\vartheta}^{23} & 2{\vartheta}^{23} &
    {\vartheta}^{03}-{\vartheta}^{13} &
    -({\vartheta}^{02}-{\vartheta}^{12}) \cr \bullet &
    2{\vartheta}^{23} & {\vartheta}^{03}-{\vartheta}^{13} &
    -({\vartheta}^{02}-{\vartheta}^{12})\cr \bullet & \bullet & 0 & 0
    \cr \bullet & \bullet & \bullet & 0 },
\end{equation}

\begin{equation}
^{(3)}Z^{\alpha\beta} = 0\,,
\end{equation}

\begin{equation}
^{(4)}Z^{\alpha\beta} = 0\,,
\end{equation}

\begin{eqnarray}
  ^{(5)}Z^{\alpha\beta} & = &
  \frac{ML_{0}r^2(r^2-3a^2{\cos^2\theta})}{2{\Delta}{\Sigma}^3}\pmatrix{
    2{\vartheta}^{01} & 0 & 0 & -{\vartheta}^{13} \cr
    \bullet & 2{\vartheta}^{01} & {\vartheta}^{02} &
    {\vartheta}^{03}\cr \bullet & \bullet & 0 & 0 \cr \bullet &
    \bullet & \bullet & 0 }\cr & & \cr & & \cr
  & & +
  \frac{ML_{0}a^2{\cos^2\theta}(3r^2-a^2{\cos^2\theta})}
  {2{\Delta}{\Sigma}^3}\pmatrix{
    0 & 2{\vartheta}^{01} & {\vartheta}^{02}-{\vartheta}^{12} & 
    {\vartheta}^{03} \cr
    \bullet & 0 & {\vartheta}^{12} & -{\vartheta}^{13}\cr \bullet &
    \bullet & 0 & 0 \cr \bullet & \bullet & \bullet & 0 }\cr & & \cr &
  & \cr
  & & +
  \frac{L_{0}a^2r^2{\sin\theta}{\cos\theta}}{{\Sigma}^3}
  \sqrt{\frac{F}{{\Delta}}} \pmatrix{-2{\vartheta}^{02} &
    -({\vartheta}^{02}+{\vartheta}^{12}) & -{\vartheta}^{01} &
    {\vartheta}^{23} \cr \bullet & -2{\vartheta}^{12} &
    -{\vartheta}^{01} & {\vartheta}^{23} \cr \bullet & \bullet &
    -2({\vartheta}^{02}-{\vartheta}^{12}) &
    -({\vartheta}^{03}-{\vartheta}^{13}) \cr \bullet & \bullet &
    \bullet & 0 }\cr & & \cr
  & & +
  \frac{L_{0}a^3r{\sin\theta}{\cos^2\theta}}{{\Sigma}^3}
  \sqrt{\frac{F}{{\Delta}}} \pmatrix{-2{\vartheta}^{03} &
    -({\vartheta}^{03}+{\vartheta}^{13}) & -{\vartheta}^{23} &
    -{\vartheta}^{01} \cr \bullet & -2{\vartheta}^{13} &
    -{\vartheta}^{23} & {\vartheta}^{01} \cr \bullet & \bullet & 0 &
    -({\vartheta}^{02}-{\vartheta}^{12})\cr \bullet & \bullet &
    \bullet & -2({\vartheta}^{03}-{\vartheta}^{13})  }\cr & & \cr & &
  + \frac{L_{0}a^2r{\sin^2\theta}F}{{\Delta}{\Sigma}^2}
  \pmatrix{0 & -2{\vartheta}^{01} & 0 &
    -{\vartheta}^{03} \cr
    \bullet & 0 & -{\vartheta}^{12} &
    {\vartheta}^{03}+{\vartheta}^{13} 
    \cr \bullet
    & \bullet & 0 & 0\cr \bullet & \bullet & \bullet & 0  }.
\end{eqnarray}

\end{footnotesize}


\centerline{================}

\end{document}